\documentclass[journal]{IEEEtran}

\normalsize
\usepackage{cite}
\usepackage{amsmath,amssymb,amsfonts}
\usepackage{graphicx}
\usepackage{caption}
\usepackage{subcaption}
\usepackage{textcomp}
\usepackage{xcolor}
\def\BibTeX{{\rm B\kern-.05em{\sc i\kern-.025em b}\kern-.08em
    T\kern-.1667em\lower.7ex\hbox{E}\kern-.125emX}}

\usepackage{verbatim}
\usepackage{comment}
\usepackage{algorithm}
\usepackage{algcompatible}
\algnewcommand\algorithmicreturn{\textbf{return}}
\algnewcommand\RETURN{\STATE \algorithmicreturn}%

\def \cC{\mathcal{C}}

\newcommand{\sort}[1]{{#1}}

\makeatletter
\newcommand\fs@spaceruled{\def\@fs@cfont{\bfseries}\let\@fs@capt\floatc@ruled
  \def\@fs@pre{\vspace{1\baselineskip}\hrule height.8pt depth0pt \kern2pt}%
  \def\@fs@post{\kern2pt\hrule\relax}%
  \def\@fs@mid{\kern2pt\hrule\kern2pt}%
  \let\@fs@iftopcapt\IFtrue}
\makeatother

\makeatletter
\newenvironment{breakablealgorithm}
  {
   \begin{center}
     \refstepcounter{algorithm}
     \hrule height.8pt depth0pt \kern2pt
     \renewcommand{\caption}[2][\relax]{
       {\raggedright\textbf{\fname@algorithm~\thealgorithm} ##2\par}%
       \ifx\relax##1\relax 
         \addcontentsline{loa}{algorithm}{\protect\numberline{\thealgorithm}##2}%
       \else 
         \addcontentsline{loa}{algorithm}{\protect\numberline{\thealgorithm}##1}%
       \fi
       \kern2pt\hrule\kern2pt
     }
  }{
     \kern2pt\hrule\relax
   \end{center}
  }
\makeatother

\def \ben{\begin{eqnarray*}}
\def \een{\end{eqnarray*}}

\def \cC{\mathcal{C}} 

\def\mod{\text{mod}}
\def\demod{\text{demod}}

\def  \code-book{C}

\def  \H5G{H_\text{CA-Polar}}

\def  \wH{w_\text{H}}
\def  \wL{w_\text{L}}

\def  \F{\mathbb {F}}

\def  \R{\mathbb {R}}
\def  \N{\mathbb {N}}

\def  \Z{\mathbb {Z}}

\def \cC{\mathcal{C}}

\def \cS{\mathcal{S}}

\def\rel{\text{Rel}}

\def\soft{ \Phi }

\def\LLR{\text{LLR}}

\newcommand{\absLLR}[1]{|\LLR(Y_{#1})|}
\newcommand{\absorbLLR}[1]{L_{#1}}

\def\relest{\lambda} 
\def\intpartset{\cS}
\def\fullintpartset{\cS}
\def\partintpartset{\Psi}
\def\splitpattern{\Xi}
\def\numsplitpattern{\xi}

\begin{document}

\title{ORDERED RELIABILITY BITS GUESSING RANDOM ADDITIVE NOISE DECODING}
\author{Ken~R.~Duffy, Wei~An,
        Muriel~M\'edard,~\IEEEmembership{Fellow,~IEEE}
        \thanks{Ken R. Duffy is with Hamilton Institute, Maynooth University, Ireland email: ken.duffy@mu.ie}
\thanks{Wei An and Muriel M\'edard are with Research Laboratory of Electronics, Massachusetts Institute of Technology, Cambridge, MA 02139, USA email: wei\_an@mit.edu,~medard@mit.edu}
\thanks{This paper was presented in part at 2021 IEEE ICASSP.}
}

\maketitle

\begin{abstract}
Error correction techniques traditionally focus on the
co-design of restricted code-structures in tandem with code-specific decoders that are 
computationally efficient when decoding long codes in hardware.
Modern applications are, however, driving 
demand for ultra-reliable low-latency communications (URLLC), rekindling interest 
in the performance of shorter, higher-rate error correcting codes, and 
raising the possibility of revisiting universal, code-agnostic decoders.

To that end, here we introduce a soft-detection variant of Guessing Random 
Additive Noise Decoding (GRAND) called Ordered Reliability Bits GRAND that 
can accurately decode any moderate redundancy block-code. It is designed
with efficient circuit implementation in mind, and determines
accurate decodings while retaining the original hard detection
GRAND algorithm's suitability for a highly parallelized implementation in hardware.

ORBGRAND is shown to provide excellent soft decision  block error 
performance for codes of distinct classes (BCH, CA-Polar and RLC) with modest complexity,
while providing better block error rate performance than CA-SCL, a
state of the art soft detection CA-Polar decoder. 
ORBGRAND offers the possibility of an accurate, energy efficient soft detection
decoder suitable for delivering URLLC in a single hardware realization.

\end{abstract}


\section{Introduction} \label{sect:intro}

Shannon's
pioneering work \cite{Shannon48} established that the highest code-rate that a channel can support is achieved as the
code becomes long. Since 1978, however, it has been known that optimally accurate Maximum Likelihood (ML) decoding of
linear 
codes is an NP-complete problem \cite{berlekamp1978inherent}. 
Taken together, those results have driven the engineering paradigm
of co-designing significantly restricted classes of linear code-books in tandem
with code-specific decoding methods that exploit the 
code-structure to enable computationally efficient approximate-ML
decoding \cite{lin2004error} for long, high-redundancy codes. 
For example, Bose-Chaudhuri-Hocquenghem (BCH) codes with
hard detection Berlekamp-Massey decoding
\cite{berlekamp1968algebraic,massey1969shift}, 
Turbo codes with soft detection iterative decoders \cite{turbo:code},
Low Density Parity Check Codes (LDPCs) \cite{gallager1963low} with soft detection
belief propagation decoding~\cite{ldpc97,fossorier1999reduced}, and
the recently proposed CRC-Assisted Polar (CA-Polar) codes, which will be used for
all control channel communications in 5G New Radio \cite{3gpp38212}, with soft detection CRC-Assisted Successive
Cancellation List (CA-SCL) decoding~\cite{KK-CA-Polar,TV-list,llr-ca-scl,leonardon2019fast, 7114328, 6823099, 9186729, 8361464, 7471817, 9020375} or other alternatives \cite{9195233, 9613535, 9613577}.

Contemporary applications, including augmented and virtual reality,
vehicle-to-vehicle communications, 
machine-type communications, and
the Internet of Things, have driven demand for Ultra-Reliable
Low-Latency Communication (URLLC)
\cite{durisi2016toward,she2017radio,chen2018ultra,parvez2018survey,medard20205}.
As realizing these technologies requires shorter codes, the
computational complexity issues associated with long codes will be vacated in delivering URLLC, 
offering the opportunity to revisit the possibility of
creating high-accuracy near-optimal universal decoders. The development of practical
universal decoders would open 
up a massively larger palette of potential code-books that can be decoded with 
a single algorithmic instantiation, greatly reducing hardware footprint,
future-proofing devices against the introduction of new codes, and enabling the 
flexibility for each application to select the most suitable code-book.


Key to unlocking that promise is the development of algorithms
that are inherently suitable for efficient implementation in circuits.
One potential approach is the recently introduced Guessing Random Additive Noise Decoding (GRAND). 
Originally established for hard decision demodulation systems \cite{Duffy18,kM:grand},
GRAND provides ML decodings for any moderate redundancy block-code
construction. It does so by sequentially removing putative noise-effects, ordered
from most likely to least likely based on a statistical channel model, from the demodulated
received sequence and querying if what remains is in the code-book. The first instance where a code-book
member is found is the decoding. Pseudo-code for GRAND can be found in Algorithm \ref{fig:grand_alg}.

\floatstyle{spaceruled}
\restylefloat{algorithm}

\begin{breakablealgorithm}
\caption{Guessing Random Additive Noise Decoding. Inputs:
a demodulated channel output $y^n = (y_1, y_2, \ldots, y_n)$; a code-book membership
function such that $\code-book(y^n)=1$ if and only if $y^n$ is in
the code-book; and optional statistical noise characteristics or soft
information, $\soft$. Output: decoded element $c^{n,*}$.}
\label{fig:grand_alg}
\begin{algorithmic}
\STATE {\bf Inputs}: Code-book membership function $C: \{0,1\}^n \mapsto \{0,1\}$; demodulated bits $y^n$; optional information $\Phi$.
\STATE {\bf Output}: Decoding $c^{n,*}$.
\STATE $d\leftarrow 0$.
\WHILE {$d=0$}
    \STATE $z^n\leftarrow$ next most likely binary noise effect sequence (which may depend on $\Phi$)
    \IF{$\code-book(y^n\ominus z^n) = 1$}
        \STATE  $c^{n,*}\leftarrow y^n\ominus z^n$
        \STATE  $d\leftarrow 1$
    \ENDIF
\ENDWHILE
\STATE{\bf return} $c^{n,*}$
\end{algorithmic}
\hrule

\end{breakablealgorithm}

Consider an arbitrary code of rate $R=k/n$ 
consisting of $M=2^k$ binary
strings of length $n$, $\cC_n=\{c^{n,1},\ldots ,c^{n,M}\}$. With $c^n=(c_1, c_2, \ldots, c_n) \in\{0,1\}^n$ being
a transmitted code-word, $y^n = (y_1, y_2, \ldots, y_n)$ denoting the hard decision demodulation, 
$Z^n = (Z_1, Z_2, \ldots, Z_n)$ denoting an independent binary additive noise-effect on the binary sequence, and $\oplus$ denoting addition
in $\F_2$, we have $y^n = c^n \oplus Z^n$.
A maximum likelihood decoding satisfies
\begin{align*}
    c^{n,*} &= \arg\max_{i\in\{1,\ldots,M\}} P(y^n|c^{n,i}) \\
    & = \arg\max_{i\in\{1,\ldots,M\}} P\left(Z^n = y^n\oplus c^{n,i}\right).
\end{align*}
Even for short codes, 
brute force identification of such a $c^{n,*}$ is not possible as it requires
$M=2^k$ computations for each decoding. 

By rank-ordering putative noise-effects in decreasing order of likelihood and
breaking ties arbitrarily, i.e. determining the sequences $\{z^{n,i}\in\{0,1\}^n\}$ such that
$P\left(Z^n = z^{n,i}\right) \geq P\left(Z^n = z^{n,j}\right)$
for all $i<j$, subtracting them from the demodulated received sequence in that order
and querying if what remains, $y^n\oplus z^{n,i}$, is in the code-book, the first such $z^{n,*}$ 
is an ML decoding so long as noise-effects are queried in decreasing order
of likelihood, even for channels with memory in
the absence of interleaving \cite{kM:grand, grand-mo,An2022TCOM}. 

For a code where $k$ information bits are transformed into $n$ coded bits, all GRAND algorithms identify an erroneous decoding after approximately geometrically distributed number of code-book queries with mean $2^{n-k}$ \cite[Theorem 2]{kM:grand} and correctly decode if they identify a code-word beforehand. Consequently,
an upper bound on the complexity of all GRAND algorithms is determined by the number of redundant bits rather than the code length or rate directly, making them suitable for decoding any moderate redundancy code of any length. 

The simplicity of GRAND's hard detection operation and the evident parallelizability of its code-book
queries have already resulted in the proposal \cite{abbas2020, abbas2021high-MO} 
and realization \cite{Riaz21} of efficient circuit implementations. The VLSI designs in \cite{abbas2020, abbas2021high-MO} focus on maximizing throughput and
minimizing worst-case latency by parallelization. The taped-out realization \cite{Riaz21} provides a universal 128-bit hard decoder  with class-leading measurements of precision, latency and energy per bit.

GRAND algorithms have two core components: a code-book membership checker and
a sequential putative noise-effect sequence generator. The former is common to all variants. 
If the code is unstructured and stored in a dictionary, a code-book query
corresponds to a tree-search with a complexity that is logarithmic
in the code-length. If the code is a Cyclic Redundancy Check (CRC)
code, which is typically only used for error detection, checking for
code-book membership requires a simple polynomial calculation.
If the code is linear in any finite field, code-book membership can be
determined by a matrix multiplication and comparison. Instead it is the putative
noise-effect sequence generator that differs with each variant in light
of statistical or per-realization information on channel characteristics.

Incorporating soft detection information into
decoding decisions is known to significantly improve accuracy
\cite{Coo88,KNH97,GS99}. Doing so requires that additional quantized soft
information be passed from the receiver to the decoder and,
for GRAND, the development of an appropriate noise-effect pattern generator that can 
accurately and efficiently create noise-effect sequences in  order of decreasing
likelihood in light of that soft information.

Symbol Reliability GRAND (SRGRAND) \cite{Duffy19a,Duffy21}
is a variant that avails of the most limited quantized soft information where
one additional bit tags each demodulated symbol as being reliably or unreliably
received. SRGRAND retains the desirable
parallelizability of the original algorithm, is readily implementable 
in hardware, and provides
a $0.5-0.75$ dB gain over hard-detection GRAND \cite{Duffy21}.
At the other extreme, Soft GRAND (SGRAND) \cite{solomon20} is a
variant that uses real-valued soft information per demodulated bit to
build a dedicated noise-effect query order for each received signal.
Using dynamic max-heap data structures,
it is possible to create a semi-parallelizable implementation in software
and, being a true soft-ML decoder, it provides a benchmark for optimal decoding accuracy
performance. However, SGRAND's execution is
algorithmically involved and does not lend itself to hardware implementation.

Here we develop Ordered Reliability Bits GRAND (ORBGRAND), 
which bridges the gap between SRGRAND and SGRAND by obtaining the
decoding accuracy of the latter in an algorithm that is, by design,
suitable for implementation in circuits. A preliminary version of ORBGRAND
that provides near-ML performance for arbitrary length, moderate-redundancy codes and  block error rates (BLER) 
greater than $10^{-3}$
was presented at IEEE ICASSP in 2021 \cite{duffy2021ordered}. Its promise 
for a highly parallelized 
hardware realization has already resulted in
VLSI architectures being proposed \cite{abbas2021high, condo2021high, condo2021fixed} and it has been used
to investigate the suitability of both existing and non-traditionally structured codes for use in URLLC \cite{Papadopoulou21, grand-crc}. Here we explain the
rationale behind ORBGRAND's design and expand on the preliminary conference version to generate near-ML performance
for higher SNR. In the process, we describe an efficient algorithm that is suitable
for hardware implementation, and establish
performance.

The rest of this paper is organized as follows. Section \ref{sec:relwork} provides a brief overview
of practical short codes and other approaches to universal soft detection decoding. Section \ref{sect:theOrb} presents 
the rationale behind ORBGRAND and its practical implementation, which 
leads to the basic and full versions of ORBGRAND. Performance evaluation results and computational complexity analysis
that demonstrate ORBGRAND's effectiveness are presented in Section \ref{sect:simulation}.  
Section \ref{sect:summary} closes with final remarks.

\section{Related work} \label{sec:relwork}
In the quest to identify short code solutions, new low-latency applications have placed renewed focus on conventional codes \cite{short_fec, bch_m2m, fec_wireless} such as Reed-Solomon Codes\cite{ReedSolo} and BCH codes\cite{journals/iandc/BoseR60a}. Soft detection decoders offer a non-trivial decoding performance gain over hard decoders \cite{lin2004error}, which will be especially necessary for short, high-rate codes. However, many traditional codes do not have corresponding soft decoders. Some state-of-art codes with dedicated soft decoders, such as Turbo and LDPC codes, can reach near Shannon-capacity performance with long codes, but their performance degrades when used with short, high-rate codes.  

Notably, Polar codes, which were the first non-random codes that were mathematically established to be capacity-achieving~\cite{Arikan09}, have received significant attention. Owing to their poor performance at practical block-lengths \cite{arikan2009performance,pfister2014brief,sarkis2015fast}, however, they have not been adopted on their own. Instead, a concatenated design has been proposed where a CRC is first added to the data, which is then Polar coded, resulting in CA-Polar codes. These codes are usually decoded with a list decoding approach where a collection of candidate Polar code-words is first determined, and then a code-word that satisfies the CRC is selected~\cite{KK-CA-Polar,TV-list,llr-ca-scl,leonardon2019fast}. As they can be constructed at short block-lengths and have an efficient soft detection decoder, CA-Polar codes have been adopted for use for all control channel communications in the 5G New Radio standard \cite{3gpp38212}. Considered as a single code, a CA-Polar code is itself a linear code, albeit one that has no dedicated decoder. As a result, GRAND algorithms have previously established that there is additional performance left to be squeezed of out of them \cite{grand-crc}.

An alternate approach to designing code-specific decoders is to instead develop a universal decoder. 
One class of soft detection decoders that can decode any binary linear code, which works on a list-decoding principle,
has been substantially investigated \cite{dorsch1974decoding,FL95,gazelle1997reliability,VF04,wu2006soft,baldi2016use,Yue21order}.  In 
Ordered Statistics Decoding (OSD), rather than compute the
conditional likelihood of the received signal for all members of
the code-book, instead the computation is done for a restricted list of candidate code-words
that is hoped to contain the transmitted one. The algorithm permutes
the columns of the parity check matrix in a manner that depends on
the received signal reliability and Gaussian elimination is then performed to rewrite
the generator matrix in systematic format, subject to checks that ensure a basis
is identified, so that the systematic element of the code is based on the most reliable bits. 
Treating the code as a hash, a candidate list of code-words is determined by placing a ball of fixed
Hamming distance around the reliable bits, and completing them with the hash. Transforming elements of
this list back into the original basis, maximum likelihood decoding is performed on the restricted list. To achieve approximate-ML decoding performance, multiple stages of reprocessing are required, making it a challenge to implement 
efficiently in hardware, especially for high throughput designs \cite{Papadopoulou21} or low power applications.

ORBGRAND inherits GRAND's potential for a highly parallelized implementation suitable for either high throughput applications or ultra-low  power for use in battery-operated devices. Leaving the code-book checker unchanged, core to ORBGRAND is a new noise-effect pattern generator that incorporates per-realization soft information in a manner that lends itself to efficient hardware implementation, as explained in the following sections.

\section{ORBGRAND} \label{sect:theOrb}

We first introduce the principle behind ORBGRAND's design, before
explaining how the basic and full variants are implemented. 

\subsection{ORBGRAND Principles} \label{sub:principle}

Using ``$\mod$'' and ``$\demod$'' as short-hand for modulation and de-modulation respectively,
an $n$-bit binary block code-word $c^n\in\{0,1\}^n$,
is modulated to $\mod(c^n)\in\{-1,1\}^n$ by $\mod(c_i) = 2c_i-1$, transmitted and impacted by independent continuous additive noise, $N^n\in \R^n$, resulting in a random
received signal $Y^n = \mod(c^n)+N^n$, from which the hard decision sequence $y^n = \demod(Y^n)$, an estimate of
$c^n$, is obtained. The noise effect is the difference between what the transmitted binary codeword 
and the demodulated received signal, $Z^n = c^n \ominus y^n$. All GRAND algorithms make queries
to identify the noise effect, $Z^n$, rather than the original continuous noise on the channel $N^n$.
With $f_{Y|C}$ being the probability density function of $Y$ given $C$, the log-likelihood ratio defined as
\begin{align*}
    \LLR(Y_i) = \log {\frac{f_{Y|C}(Y_i| 1)}{f_{Y|C}(Y_i|0)}}, 
\end{align*}
the hard detection $y_i$ is obtained from $Y_i$ by $y_i = (\text{sign}(\LLR(Y_i))+1)/2$,
and 
$\absLLR{i}$
is referred to as the reliability of $y_i$.

While there are many ways to quantitatively capture the soft information in $Y^n$, 
for ORBGRAND it is instructive to first represent it as a sequence, 
$B^n=(B_1,B_2,\ldots ,B_n)$, where $B_i$ is the {\it a posteriori}
likelihood that the hard decision bit $y_i$ is in error, which 
can be expressed in terms of the bit reliabilities as
\begin{align} \label{eq:B_llr}
B_i = \frac{e^{-\absLLR{i}}}{1 + e^{-\absLLR{i}}} \in[0,1/2],
\end{align}
where $B_i$ is monotonically decreasing with $\absLLR{i}$. From $B^n$ we can evaluate the {\it a posteriori} likelihood of a
binary noise-effect sequence $z^n$,
\begin{align*} 
& P(Z^n=z^n) \\
&=\prod_{i:z_i=0} (1-B_i) \prod_{i:z_i=1} B_i \nonumber = \prod_{i=1}^n (1-B_i) \prod_{i:z_i=1} \frac{B_i}{1-B_i} \nonumber  \\
&\propto \prod_{i:z_i=1} \frac{B_i}{1-B_i} = \exp\left(-\sum_{i=1}^n \absLLR{i}z_i\right).
\end{align*}
Therefore, up to a constant shared by all sequences, the likelihood of a putative noise effect sequence $z^n$ is determined by the sum of the reliabilities of hard-detected bits being flipped, 
$\rel(z^n)= \sum_{i=1}^n \absLLR{i} z_i$.
To rank order putative noise sequences, $z^n$, in decreasing likelihood, it is, therefore, sufficient to rank order them by increasing reliability sum, $\rel(z^n)$.

If no soft information is available, by defining $\absLLR{i}$ to be an arbitrary positive constant for all $i$, $\rel(z^n)$ is proportional to the Hamming Weight of $z^n$, $\wH(z^n)=\sum_{i=1}^n z_i$. In this case, putative noise sequences would be rank ordered in increasing Hamming weight, as used in the original hard detection GRAND for a binary symmetric channel. SRGRAND filters $\absLLR{i}$, setting it to be $+\infty$ if it is above a threshold and to a positive constant if below that threshold, resulting in putative noise sequences being be rank ordered in increasing Hamming weight within the masked region of finite reliability bits. Armed with $\{\absLLR{i}:i\in\{1,\ldots,n\}\}$, true soft ML decoding is achieved by SGRAND using a dynamic algorithm that recursively generates a max-heap for each set of reliabilities to generate $z^n$ with increasing $\rel(z^n)$. Our goal with ORBGRAND is to obtain comparable performance with an algorithm that is amenable to efficient implementation by design. 

For notational simplicity, we shall assume that the reliabilities, $\{\absLLR{i}:i\in\{1,\ldots,n\}\}$, happen to be received in increasing order of bit position, so that $\absLLR{i}\leq\absLLR{j}$ for $i\leq j$. In practice, for each received block we sort the reliabilities and store the permutation, $\pi^n=(\pi_1,\ldots,\pi_n)$, such that $\pi_i$ records the received order index of the $i^\text{th}$ least reliable bit. The permutation $\pi^n$ enables us to map all considerations back to the original order that the bits were received in. 

\begin{figure}[htbp]
\centerline{\includegraphics[width=0.45\textwidth]{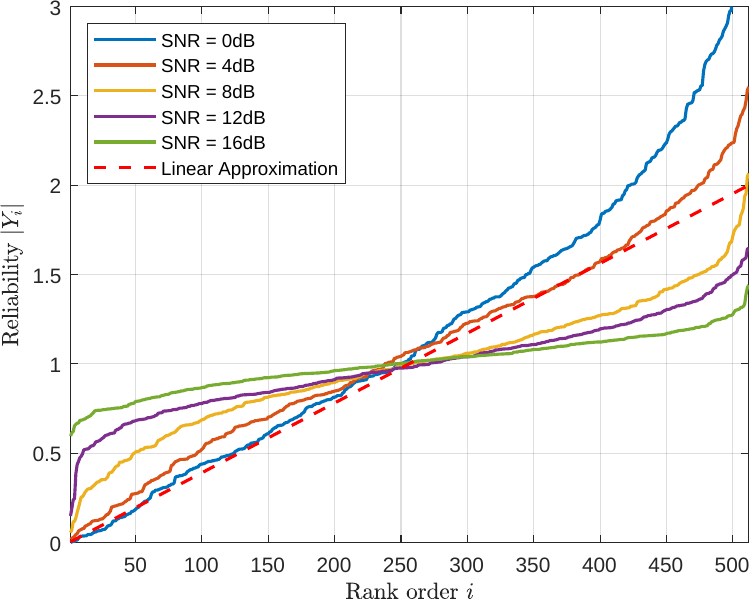}}
\caption{Samples of ordered reliability of 512-bit sequences in AWGN channel with given SNRs.}
\label{fig:ordered_rlbl}
\end{figure}

The core of the approach underlying ORBGRAND is the development of statistical models of the non-decreasing sequence $\{\absLLR{i}:i\in\{1,\ldots,n\}\}$ that are accurate, robust, and lead to computational efficient algorithms for generating rank ordered putative noise sequences. The approach can be most readily understood with the example of a channel using BPSK modulation that is subject to Additive White Gaussian Noise (AWGN), where
$\LLR(Y) \propto Y$. As constants of proportionality will prove to have no impact on ORBGRAND's order, from here on we will refer to
$\absorbLLR{i}=|Y_i|$ as  the reliability of the $i$-th bit.
Sample  rank ordered reliability values $\{\absorbLLR{i}:i\in\{1,\ldots,n\}\}$ are plotted in Fig. \ref{fig:ordered_rlbl} for various SNRs.
At lower SNR, the reliability curve is near linear with a zero intercept, while for high SNR the intercept is non-zero and there is notable curvature, particular for the least reliable bits, which are most significant for generating an accurate query order. 
Different levels of approximation to the reliability curve lead to distinct decoding complexity and performance, as will be explored in the following sections.

\subsection{Basic ORBGRAND - The Low SNR Model}
\label{sub:basicorb}

The simplest statistical model, $\relest^n=(\relest_1,\ldots,\relest_n)$, for the reliability curve is a line through the origin with 
slope $\beta>0$,
\begin{align}
    \relest_i = \beta\, i \text{,~for~} i=1, 2, \ldots, n.
    \label{eq:llr2i}
\end{align}
This model is illustrated by the dashed line in Fig. \ref{fig:ordered_rlbl},
where it can be seen to provide a good approximation at lower SNR. 
For the zero-intercept linear model,
\begin{align} \label{eq:lw1}
\rel(\sort{z}^n) \approx \sum_{i:\sort{z}_i=1} \relest_i = \beta \sum_{i=1}^n i z_i = \beta
\wL(\sort{z}^n),  
\end{align}
where we define  the sum of the positions that are flipped,
\begin{align} \label{eq:lw}
\wL(z^n)=\sum_{i=1}^n i z_i,
\end{align}
to be the {\it Logistic Weight} of the binary sequence $z^n$. 
Thus, in this model the likelihoods of putative noise effect sequences are ordered in increasing logistic weight and hence the value of $\beta$ need not be estimated.

\begin{figure}[ht]
\begin{center}
\includegraphics[width=0.38\textwidth]{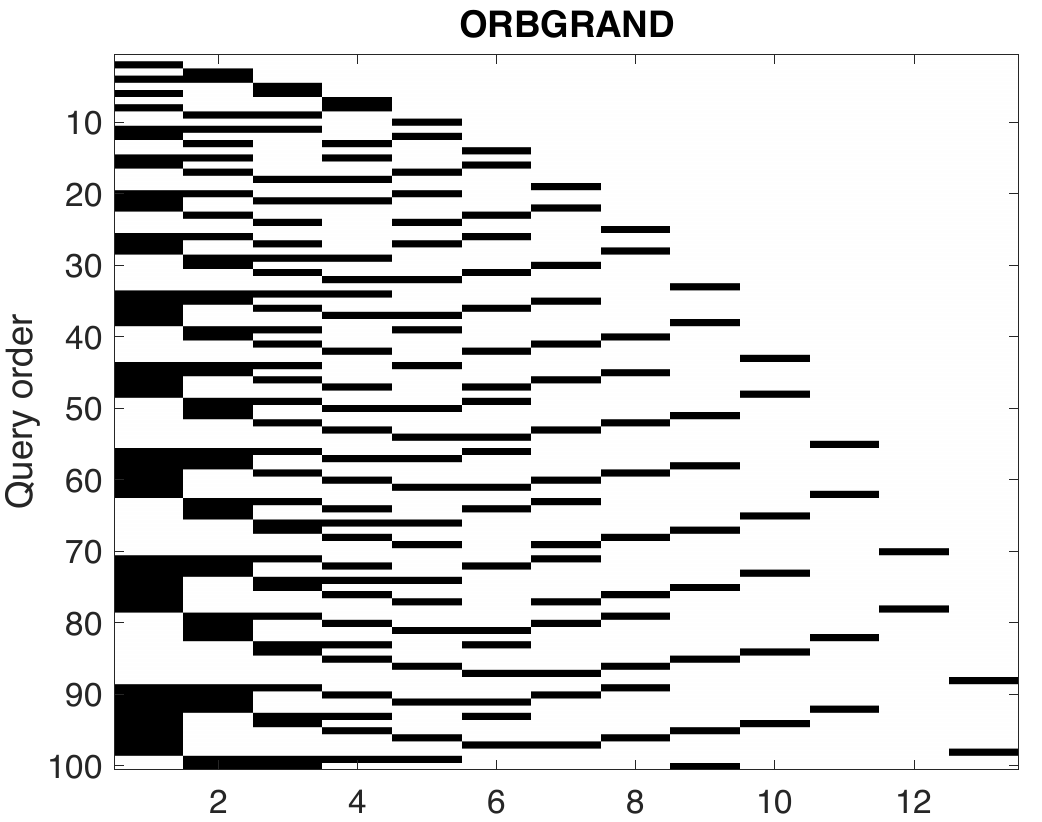}
\end{center}
\caption{First 100 ORBGRAND noise effect queries where bit positions are
in increasing order of Logistic Weight. Each
row is a noise sequence with white being no bit
flip and black corresponding to a bit flip.}
\label{fig:basic_orb_pat}
\end{figure}

Consequently, for any $\beta$ the first putative error sequence always corresponds to no bits being flipped, which has $\wL=0$. The second query corresponds to $\sort{z}^n$ with the least reliable bit flipped, having $\wL=1$. The third corresponds to only the second least reliable bit of $\sort{z}^n$ flipped, which has $\wL=2$.
The next query is either the noise-effect where only the third least reliable bit is flipped or the one where the least reliable and second least reliable bits are both flipped, both having $\wL=3$, with the tie broken arbitrarily. The ordering proceeds in that fashion as illustrated in Fig. \ref{fig:basic_orb_pat}, which describes the noise-effect sequence generator in basic ORBGRAND \cite{duffy2021ordered}. Thus, for its operation, ORBGRAND based on this statistical model only requires the permutation recording the positions of the rank ordered reliabilities of the received bits, $\pi^n$, from which the algorithm proceeds deterministically. 

What remains to do for basic ORBGRAND is to develop an efficient algorithm that sequentially generates putative noise sequences in terms of increasing logistic weight. Noting that for a binary string of length $n$ the maximum logistic weight is achieved by the sequence of all $1$s giving $\wL(1,\ldots,1)=n(n+1)/2$, to achieve the goal we must be able to identify all allowable noise-effect sequences for each  logistic weight $W\in\{0,\ldots, n(n+1)/2\}$,
\begin{align} \label{eq:znwl}
\intpartset_W = \left\{\sort{z}^n \in\{0,1\}^n: \wL(\sort{z}^n) =W\right\}.
\end{align}
That objective can be fractionated by conditioning on the Hamming weight, $w$,
of the sequences, giving
\begin{align}
&\intpartset_W = \label{eq:fractionation}\\
& \bigcup_{w=1}^{\lfloor (\sqrt{1+8W}-1)/2\rfloor} \left\{\sort{z}^n \in\{0,1\}^n: \wH(\sort{z}^n)=w, \wL(\sort{z}^n) =W\right\},
\nonumber
\end{align}
where the upper-bound on the union stems from the fact that if the Hamming weight of $\sort{z}^n$ is $w$, the smallest
logistic weight that $\sort{z}^n$ can have is from the sequence with flipped bits in the first $w$ positions of $\sort{z}^n$, giving a logistic weight of $\wL(z^n) = w(w+1)/2 \leq W$.

Consider a single set in the union in Eq. \eqref{eq:fractionation} for Hamming weight $w$. Determining
\begin{align*}
\left\{\sort{z}^n \in\{0,1\}^n: \wH(\sort{z}^n)=w, \wL(\sort{z}^n) =W\right\}
\end{align*}
is equivalent to finding all integer-valued vectors of length $w$ satisfying
\begin{align}
\left\{v^w \in\N^w : 1\leq v_1<\ldots <v_w\leq n, \sum_{i=1}^w v_i =W\right\},
\label{eq:intpart}
\end{align}
where $v^w$ contains the indices of the flipped bits in $\sort{z}^n$, which amounts
to finding all integer partitions of $W$ of size $w$ with non-repeating positive parts subject to a
maximum value of $n$. By setting 
\begin{align}
v_i = i + u_i, \text{~for~}i=1, 2, \ldots, w,
\label{eq:vu}
\end{align}
it is possible to reformulate the set in Eq. \eqref{eq:intpart} in one final way in terms of the $u_i$, as the integer partitions of $W'=W-w(w+1)/2$ into $w$ not-necessarily distinct, non-negative parts  no larger than $n'=n-w$. 
That is, determining all the elements in the set Eq. \eqref{eq:intpart}, is equivalent to finding all integer vectors $u^w$ such that
\begin{equation}
\left\{ u^w \in\Z_+^w : 0 \leq u_1 \leq u_2 \leq \ldots \leq u_w\leq n', \sum_{i=1}^w u_i =W' \right\}.
\label{eq:intpart2}
\end{equation}
Here we introduce an efficient algorithm for determining all sequences that are in the partition, which 
is suitable for implementation
in hardware. It will form an essential component of the full ORBGRAND, which uses a more sophisticated
model than described in Eq. \eqref{eq:llr2i}.

\subsection{Integer Partition Pattern Generator}
\label{subsec:land}

\begin{figure}[ht]
\begin{center}
\includegraphics[width=0.38\textwidth]{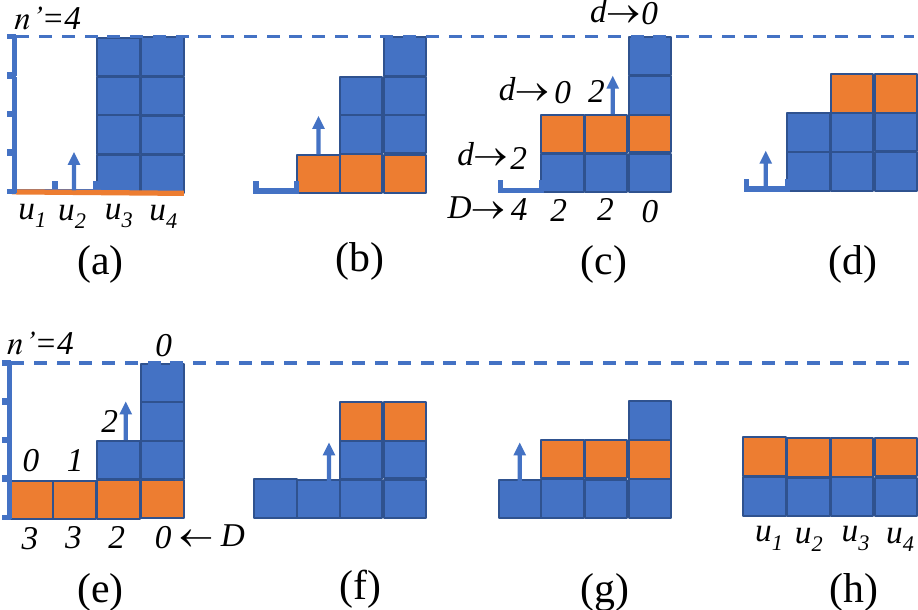}
\end{center}
\caption{Procedure for partitioning $W'=8$ into $w=4$ non-negative, non-decreasing parts, each no larger than $n'=4$. The upward arrow indicates the corresponding part is to be increased by 1 in the next step. (c) and (e) mark the values of $d(i)$ and $D(i)$ for $1\leq i \leq 4$.}
\label{fig:part_diag}
\end{figure}

Integer partitions can be represented by diagrams \cite{Cameron2001} as illustrated in Fig. \ref{fig:part_diag}, where each column represents an integer part with its value, $u_i$, equaling to the number of cells in the column and the total number of cells in the diagram equaling the integer to be partitioned $\sum_i u_i$. Here we use a mirror image of a Ferrers Diagram, where the parts are listed in the increasing order to assist in the description of the algorithm. 

A function $d:\{1,\ldots, w\}\mapsto\{0,\ldots,W'\}$ records the ``drop'' between adjacent $i$, $i+1$ parts
\begin{align*} 
d(i) = \begin{cases}
 0 & \text{ if } i=w\\
 u_{i+1}-u_i & \text{ if } i\in\{1,\ldots,w-1\}\\
\end{cases}
\end{align*}
from which the accumulated drop function is defined by
$D(i) = \sum_{j=i}^wd(j)$, where $D(1)$ records the total drop in the integer partition $u^w$. Examples of $d(i)$ and $D(i)$ are shown in Fig. \ref{fig:part_diag} (c) and (e). Fig. \ref{fig:part_diag} (a) represents an extreme case in which the minimum number of non-zero integer parts is achieved by pushing cells to the right with part values maximized. Another extreme case is that cells are spread to maximum number of parts achieving the minimum number of rows, or equivalently, satisfying $D(1) \leq 1$, as illustrated in Fig. \ref{fig:part_diag} (h). All partitions for the setting of $W'=8$, $w=4$ and $n'=4$ are obtained in the migration procedure from (a) to (h), which can be accomplished with the Landslide algorithm presented in Algorithm \ref{alg:land}.

The algorithm heavily relies on the Build-mountain routine, in which a partial partition is performed to push unallocated cells to the right-most parts, akin to building the steepest, highest mountain allowable on the right side of the diagram. For example, in Fig. \ref{fig:part_diag} (e), $u_1=1$ is determined from step (d), the remaining 7 cells are to be assigned to $u_2$, $u_3$ and $u_4$. The assignment can be accomplished by first making $u_2$, $u_3$ and $u_4$ identical to $u_1=1$, and then assigning the remaining 4 cells to the right-most parts, with $u_4$ maximized and $u_3$ increased by 1.

In general, when the values of $\{u_1,u_2, \ldots, u_k \}$ have been specified, the allocation of the remaining cells to $\{u_{k+1},u_{k+2}, \ldots, u_w \}$, or the Build-mountain routine, is carried out as follows:
\begin{enumerate}
    \item $u_i \gets u_k$, for $k+1 \leq i\leq w$
    \item $W'' \gets W'-\sum_{i=1}^{w}u_i$
    \item Obtain $q$ and $r$ such that $W'' = q(n'-u_k) + r$
    \item if $q \neq 0$, $u_i \gets n'$, for $w-q+1 \leq i \leq w$
    \item $u_{w-q} \gets u_{w-q} + r$
\end{enumerate}
The initial partition in Fig. \ref{fig:part_diag} (a) is obtained with the same method by simply assuming a dummy part $u_0=0$. With the Build-mountain routine explained, the Landslide algorithm is described as in Algorithm \ref{alg:land}.

\floatstyle{spaceruled}
\restylefloat{algorithm}

\begin{breakablealgorithm}
\caption{The Landslide Algorithm}
\label{alg:land}
\begin{flushleft}
        \textbf{Input:} $W'$, $w$, $n'$  \newline
        \textbf{Output:} $\{u^{w,j}, j=1, 2, \ldots\}$
\end{flushleft}
\begin{algorithmic}[1]

\STATE Build-mountain for initial partition
\STATE $j \gets 1$
\STATE $u^{w, j} \gets u^w$
\STATE Update $D(i)$ for $1 \leq i \leq w$

\WHILE{$D(1) \geq 2$}

\STATE Locate the largest $k$ such that $D(k) \geq 2$
\STATE $u_k \gets u_k+1$

\STATE Build-mountain from $u_k$
\STATE Update $D(i)$ for $1 \leq i \leq w$

\STATE $j \gets j+1$
\STATE $u^{w, j} \gets u^w$

\ENDWHILE

\STATE Return $\left\{u^{w,1}, u^{w,2}, u^{w,3}, \ldots \right\}$

\end{algorithmic}
\end{breakablealgorithm}

Using the same example from Fig. \ref{fig:part_diag}, the procedure of the Landslide algorithm is illustrated in Algorithm \ref{alg:land}, along with the mapping from partition $u^{w,i}$ to $v^{w,i}$ according to Eq. \eqref{eq:vu}. The diagram indicates the potential for efficient implementation of the Landslide algorithm. While one routine generates partitions for one Hamming weight $w$ at a time, multiple parallel routines can generate partitions for different Hamming weights, providing sufficient noise-effect sequences for highly-parallelized code-book checking. 

\begin{figure}[htbp]
\centerline{\includegraphics[width=0.49\textwidth]{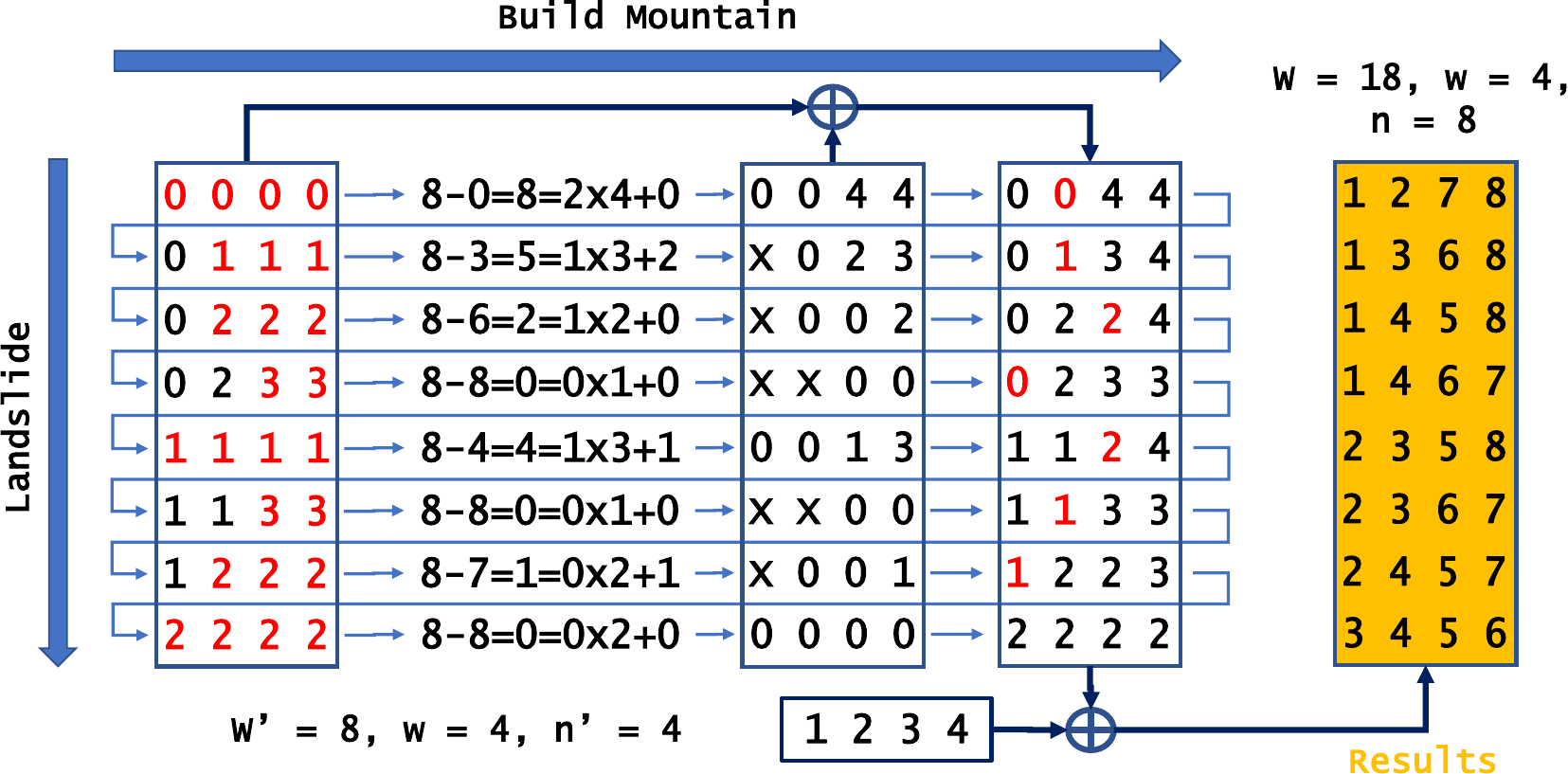}}
\caption{The Landslide algorithm is applied to achieve partitioning $W=18$ into $w=4$ distinguished parts with maximum value of $n=8$; The partition problem is first converted to partitioning $W'=8$ into $w=4$ repeatable parts with maximum value of $n'=4$; Mapping from the latter partition to the former one is simply achieved by adding 1, 2, 3, 4 individually}
\label{fig:land}
\end{figure}

\subsection{The full ORBGRAND Algorithm} \label{sub:fullORB}

The zero-intercept, linear statistical model for rank-ordered bit reliabilities that underpins basic ORBGRAND in Eq. \eqref{eq:llr2i} requires no input beyond a rank ordering of received hard-detection bits by increasing reliability and provides a good approximation to the reliability curve in low SNR conditions. It is, however, evidently a poor description at higher SNR in Fig. \ref{fig:ordered_rlbl}. That mismatch results in basic ORBGRAND's query order diverging from true likelihood order at higher SNR, with corresponding performance loss. By expanding the statistical model used to describe the reliability data to a piece-wise linear one for full ORBGRAND, we retain the algorithmic efficiencies of generating integer partition sequences while improving block error rate performance at higher SNR. 

\begin{figure}[htbp]
\centerline{\includegraphics[width=0.38\textwidth]{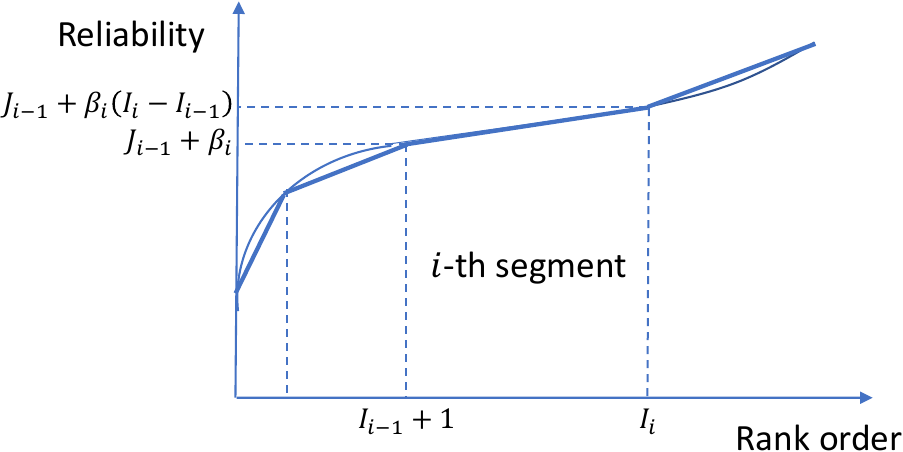}}
\caption{Full piece-wise linear statistical model to the ordered reliability curve used in ORBGRAND, with the start and end indices indicated for the $i$-th segment.}
\label{fig:single_line}
\end{figure}

As illustrated in Fig. \ref{fig:single_line}, with  $I_0=0$ and $I_m=n$,
the $m$-segment statistical model curve is represented as
\begin{align} \label{eq:relest_seg}
\relest_j = J_{i-1}+\beta_i(j-I_{i-1}) \text{,~for~} I_{i-1}<j \leq I_i,
\end{align}
where $1 \leq i \leq m$ is the segment index. The anchor indices $\{I_i: i\in\{0, 1, \ldots, m\}\}$ define 
the domain of each segment, while $J_{i-1} \in\Z$ and $\beta_i\in\N$, respectively, determine the initial value and slope of the $i$-th segment. That $J_{i-1}$ and $\beta_i$ are restricted to being integers is crucial to enabling efficient algorithmic implementation producing rank ordered putative noise sequences, and results will demonstrate that no loss in performance results from this constraint. The model used for basic ORBGRAND, Eq. \eqref{eq:llr2i}, is a special case of Eq. \eqref{eq:relest_seg} with $m=1$, $I_1=n$ and $J_0=0$. 

The approximate reliability sum of $z^n$, namely the reliability weight, based on the full model is then
\begin{align*} 
\rel(\sort{z}^n) &
\approx\sum_{i:\sort{z}_i=1}\relest_{i} 
= \sum_{i=1}^n \relest_{i}z_i 
= \sum_{i=1}^m\sum_{j=I_{i-1}+1}^{I_i} \relest_{j}z_j \\
&= \sum_{i=1}^m J_{i-1}\wH{(z_{I_{i-1}+1}, \ldots z_{I_i})} \\
&+\sum_{i=1}^m \beta_i  \wL{(z_{I_{i-1}+1}, \ldots z_{I_i})}
\in\Z_+,
\end{align*}
and the likelihood of noise effect sequences decreases with increasing reliability weight.
With this new approximation, the set of noise-effect sequences for a weight of $W$ becomes
\begin{align} 
\fullintpartset_W
&=\left\{\sort{z}^n \in\{0,1\}^n: \sum_{i=1}^m\sum_{j=I_{i-1}+1}^{I_i} \relest_{j}\sort{z}_j =W\right\} \nonumber\\
&= \bigcup_{W^m:\sum_{i=1}^m W_i=W} \left(\partintpartset^1_{W_1} \times \partintpartset^2_{W_2} 
\times \cdots \partintpartset^m_{W_m} \right),
\label{eq:cartP}
\end{align}
where
\begin{align*} 
\partintpartset^i_{W_i}=\left\{(z_{I_{i-1}+1},\ldots,z_{I_i}): \sum_{j=I_{i-1}+1}^{I_i} \relest_{j}z_j = W_i\right\}
\end{align*}
for $i\in\{1,\ldots,m\}$ and $\times$ represents Cartesian product.
Thus, to generate all elements of $\fullintpartset_W$ in Eq. \eqref{eq:cartP}, we identify the
set of all possible splitting patterns of $W$, denoted by
\begin{align} \label{eq:Pw}
\splitpattern_W = \left\{W^m \in\Z_+^m : \sum_{i=1}^{m}W_i = W \right\},
\end{align}
using Algorithm \ref{alg:split}, explained later.

For a given $W^m=(W_1,\ldots,W_m) \in \splitpattern_W$, consider the generation of the partial sequence set $\partintpartset^i_{W_i}$  defined in Eq. \eqref{eq:cartP}. Recalling Eq. \eqref{eq:relest_seg}, each partial sequence must satisfy
\begin{align} 
   & W_i = \sum_{j=I_{i-1}+1}^{I_i} (J_{i-1} + (j-I_{i-1})\beta_i)z_j \nonumber
    \\ & = J_{i-1}\wH{(z_{I_{i-1}+1}, \ldots z_{I_i})}+\beta_i\wL{(z_{I_{i-1}+1}, \ldots z_{I_i})}, \label{eq:W2IJ}
\end{align}
which, defining $w_i=\wH{(z_{I_{i-1}+1}, \ldots z_{I_i})}$ and  with $v_k = j_k-I_{i-1}$
being the relative indices of the flipped bits,
is equivalent to
\begin{align} \label{eq:wu_i}
\wL{(z_{I_{i-1}+1}, \ldots z_{I_i})} = \sum_{k=1}^{w_i} v_k = \frac{W_i - w_i J_{i-1}}{\beta_i}.
\end{align}
Eq. \eqref{eq:wu_i} indicates that, with the partial reliability weight $W_i$ and Hamming weight $w_i$ specified for the $i$-th segment, the partial noise-effect sequence generation reduces to the integer partition problem that is efficiently solved by the Landslide algorithm in section \ref{subsec:land}. 

Splitting a reliability weight value of $W$ into $m$ parts, as defined in Eq. \eqref{eq:Pw}, is a distinct integer partition problem, which we call the integer splitting problem for differentiation. The difference here lies in that the same group of parts with different orders are distinct splitting patterns. A common approach to finding all splitting patterns in $\splitpattern_W$ is given in Algorithm \ref{alg:split}, which starts with sweeping $W_1$ from 0 to $W$. For a given value of $W_1$, $W_2$ is swept from 0 to $W-W_1$. For each fixed $W_1$ and $W_2$, $W_3$ is swept and the nested loop reaches $W_{m-1}$. Then $W_m$ is computed as $W-\sum_{j=1}^{m-1}W_j$, ensuring the sum of all parts is $W$. The size of the set $\splitpattern_W$ obtained from the algorithm is $\numsplitpattern_W = \binom{W+m}{m-1}$.

\floatstyle{spaceruled}
\restylefloat{algorithm}

\begin{breakablealgorithm}
\caption{The Integer Splitting Algorithm}
\label{alg:split}
\begin{flushleft}
        \textbf{Input:} $W$, m  \newline
        \textbf{Output:} $\{W^{m,k} : k=1, 2, \ldots,\numsplitpattern_W \}$
\end{flushleft}
\begin{algorithmic}[1]

\STATE $k \gets 0$
\FOR {$W_1 = 0$ To $W$}
    \FOR {$W_2 = 0$ To $W-W_1$}
        \STATE \ldots\ldots (nested loops over $W_i$, $i=3, 4, \ldots, m-2$)
        \FOR {$W_{m-1} = 0$ To $W-\sum_{l=1}^{m-2}W_l$}
            \STATE $W_m \gets W-\sum_{l=1}^{m-1}W_l$
            \STATE $k \gets k+1$
            \STATE $W^{m, k} \gets \{W_1, W_2, \ldots, W_m\}$
        \ENDFOR
        \STATE \ldots\ldots (nested loops over $W_i$, $i=3, 4, \ldots, m-2$)
    \ENDFOR
\ENDFOR
\RETURN{} $\{W^{m,k} : k=1, 2, \ldots,\numsplitpattern_W\}$
\end{algorithmic}
\end{breakablealgorithm}

Eq. \eqref{eq:wu_i} indicates that the actual number of valid splitting patterns is, however, much smaller than $\numsplitpattern_W$, owning to the requirement that each element $W_i$ of a valid $W^m$ must satisfy all of:
\begin{align} \label{eq:wi_cond}
    \begin{cases}
     & W_i  = 0 \text{~or~} W_i - w_i J_{i-1} \ge \frac{(1+w_i)w_i}{2} \\
     & W_i - w_i J_{i-1} \le (I_i-I_{i-1}+1)w_i - \frac{(1+w_i)w_i}{2} \\
     & W_i - w_i J_{i-1} \text{~is divisible by~} \beta_i.
    \end{cases}
\end{align}
Therefore, any non-zero element $W_i$ in $W^m$ must be associated with a non-empty set of partial Hamming weights $\{w_i\}$, such that Eq. \eqref{eq:wi_cond} is satisfied. Otherwise $W^m$ is invalid and should be discarded. The associated set for $W_i$ can be obtained with Algorithm \ref{alg:phm}. 

\floatstyle{spaceruled}
\restylefloat{algorithm}

\begin{breakablealgorithm}
\caption{The collection algorithm for valid partial Hamming weights}
\label{alg:phm}
\begin{flushleft}
        \textbf{Input:} $W_i$, $J_{i-1}$  \newline
        \textbf{Output:} $\{ w_{i,k} : k=1, 2, \ldots \}$ or FAIL
\end{flushleft}
\begin{algorithmic}[1]

\STATE $k \gets 0$
\FOR {$w = 1$ To ${\lfloor \frac{\sqrt{1+8W_i}-1)}{2}\rfloor}$}
    \IF {$W_i, w$ and $J_{i-1}$ satisfy Eq. \eqref{eq:wi_cond}}
        \STATE $k \gets k + 1$
        \STATE $w_{i,k} \gets w$
    \ENDIF
\ENDFOR
\IF {$k$ is $0$}
    \RETURN{} FAIL
\ELSE
    \RETURN{} $\{ w_{i,k}: k=1, 2, \ldots \}$
\ENDIF
\end{algorithmic}
\end{breakablealgorithm}

The FAIL return from Algorithm 3 invalidates $W_i$ as well as the whole split pattern $W^m$. In Algorithm 2, each new value of $W_i$ is checked against Algorithm 3. A return of FAIL discard the current value of $W_i$ and force the loop to jump to the next iteration with a new value of $W_i$. Only when $W_i$ is validated, can the follow-up nested loop over $W_{i+1}$ continue. Each returned partial Hamming weights set $\{ w_{i,k} \text{,~} k=1, 2, \ldots \}$ should also be saved for the later generation of partial noise-effect sequences.

In addition to the validation from Algorithm \ref{alg:phm}, more measures are available for further reduction of the set size of $\splitpattern_W$. For example, after the initial value of 0, $W_i$ can jump to $J_{i-1}+\beta_i$ omitting all values in between. Generally, due to the small segment number $m$ in practice, the generation of splitting patterns $W^m$ has limited impact on the overall efficiency of the ORBGRAND algorithm, which is instead dominated by the efficient Landslide algorithm. 

A significant complexity reduction is, however, available if $J_{i-1}$ is divisible by $\beta_i$. In this case, $W_i$ must also be divisible by $\beta_i$ in order to have Eq. \eqref{eq:wi_cond} satisfied. This can be  achieved by sweeping $W_i$ in steps of size  $\beta_i$. Then the validation of a partial Hamming weight $w_i$ is straightforward, forsaking the need of Algorithm \ref{alg:phm}. The extra restriction on $J_{i-1}$ logically leads to a potential performance loss. As demonstrated by later simulations, the minor performance loss justifies the complexity reduction measure. 

\begin{figure}[htbp]
\centerline{\includegraphics[width=0.38\textwidth]{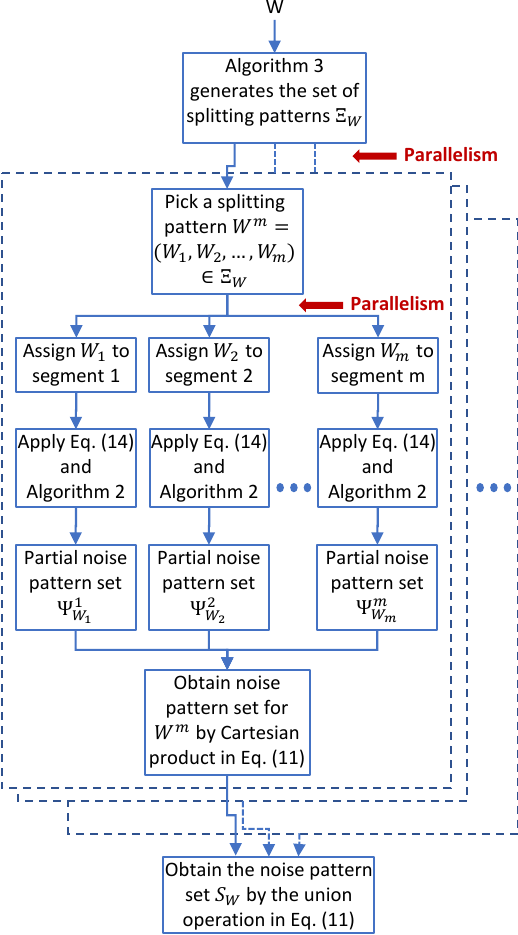}}
\caption{The procedure for generating noise sequence patterns of a given reliability weight $W$ in ORBGRAND with the potential locations for parallel implementation marked.}
\label{fig:flow_char}
\end{figure}

Given parameters of the statistical model in Eq. \eqref{eq:relest_seg}, all the components necessary to create the full ORBGRAND algorithm have been described. The likelihood order of generated noise-effect sequences is governed by the increasing value of reliability weight. For each specified weight value $W$, Algorithm \ref{alg:split} (or its optimized version) is used to generate $\splitpattern_W$, the set of valid splitting patterns. Each splitting pattern $W^m \in \splitpattern_W$ has its element (or partial reliability weight) $W_i$ assigned to the $i$-th segments. In each segment, Eq. \eqref{eq:wu_i} indicates that the Landslide algorithm can efficiently generate $\partintpartset_{W_i}$, the set of all possible partial noise-effect patterns, as defined in Eq. \eqref{eq:cartP}. The Cartesian product over partial sequence sets, as shown in Eq. \eqref{eq:cartP} is performed to create the set of noise-effect sequences for the current splitting pattern $W^m$. Finally, the union in Eq. \eqref{eq:cartP} forms $\fullintpartset_W$, the full set of noise-effect sequences for the reliability weight $W$. Parallel implementation can be achieved at several levels, such as jointly generating partial sequences for multiple segments, or concurrently generating noise-effect sequences for multiple splitting patterns. The complete procedure is presented in the flow-chart in Fig. \ref{fig:flow_char} with potential parallelisation points marked. What remains is to determine the parameters of the piece-wise linear model.

\subsection{Piece-wise linear fitting and quantization} \label{sub:fit}

Key to ORBGRAND's practical complexity is that it operates on $\relest^n=(\relest_1,\ldots,\relest_n)$, an approximation to the original rank ordered reliability curve for a given received code block $(\absorbLLR{1},\ldots, \absorbLLR{n})$. The approximation level determines the trade-off between algorithmic complexity and decoding precision. The simplest statistical model is a line through the origin, which solely requires knowledge of the rank order of the received bits by their reliability, but results in degraded performance at higher SNR scenarios. The model underlying the full ORBGRAND necessitates two stages: piece-wise linear fitting and quantization. While there are numerous approaches for either higher accuracy or lower complexity, here we introduce a method with moderate algorithmic complexity that serves as a reference design and demonstrates the robustness of ORBGRAND.

Given an independent and identically distributed set of random variables, $\{A_i:i\in\{1,\ldots,n\}\}$, drawn from a cumulative distribution $F_A$, results from the theory of Order Statistics \cite{david2004order} tell us that rank ordering from least to greatest, so that $A_{(i)}$ is the $i$-th smallest value, leads to $A_{(i)} \approx F^{-1}_A\left(i/n\right)$ for $1 \leq i \leq n$ and  large $n$. $F^{-1}_A(\cdot)$ is a monotonically increasing function and serves as the functional mean of rank ordered ensembles of observations $\{A_i:i\in\{1,\ldots,n\}\}$. 
For rank ordered reliabilities of blocks of bits received from the channel, $(\absorbLLR{1},\ldots, \absorbLLR{n})$, this serves as guidance for a fitting procedure for the statistical model.

\begin{figure}[htbp]
\centerline{\includegraphics[width=0.38\textwidth]{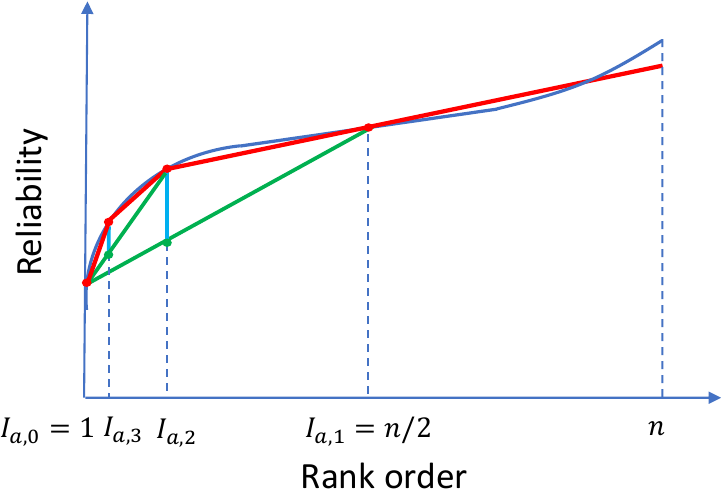}}
\caption{Identification of anchor points for segmentation of the reliability curve.}
\label{fig:dynamic}
\end{figure}

We can, therefore, use the edge point at index $I_{a,0}=1$ and the center point at index $I_{a,1}=n/2$ on $L^n$ as the initial set of anchor points from which other anchor points for segmentation can be found, as illustrated in Fig. \ref{fig:dynamic}. A straight line is drawn linking the anchor points at $I_{a,0}$ and $I_{a,1}$. The maximum vertical gap between the straight line and the reliability curve determines the location of the new anchor point with its index marked as $I_{a,2}$. New anchor points can be found between adjacent anchor points in the same way. A rule of thumb is that more points should be located in the high-curvature area near the edge. The indices of anchor points define the segmentation of the reliability curve, and the lines linking adjacent anchor points form a piece-wise linear fitting to the reliability curve. 

If the curve $L^n$ is close to a straight line between two anchor points an additional segment is unnecessary, or, a casually added segment has little impact to the performance except for some overhead in the splitting of logistic weight. The same fitting technique can be applied to the high reliability area near the right edge, however, as shown in Fig. \ref{fig:dynamic}, we choose to extend the central line to cover the area. In low SNR cases, the extended straight line by itself is a good approximation, and in high SNR cases, high reliability bits have little influence on the generation order of noise-effect sequences. The assertion has been verified with simulations. 

From Eq. \eqref{eq:relest_seg}, the piece-wise linear approximating curve is defined with three sets of non-negative integer parameters: $I_i \in\Z_+, 0 \leq i \leq m$, the indices for segmentation;
$J_i \in\Z, 0 \leq i \leq m-1$, the offset of each linear segment; and $\beta_i \in\N, 1 \leq i \leq m$, the slope of each segment. When $m+1$ anchor points on $L^n$ have been obtained, their indices are used as the segmentation indices and is denoted as $I_i, i=0, 1, 2, \ldots, m$, where $I_0=0$ and $I_m=n$. We further use the smallest slope of the fitted lines to quantize parameters, which is computed with a quantization parameter
\begin{align} \label{eq:Q}
    Q = \min \left \{ \frac{L_{I_1}-L_1}{I_1-1},\min_{i\in\{2,\ldots,m\}} \left\{ \frac{L_{I_i}-L_{I_{i-1}}}{I_i-I_{i-1}} \right\} \right\}
\end{align}
where the slope of the first segment has a different form because of the absence of $L_0$. The quantized parameters of lines are then computed as,
\begin{equation}  \label{eq:quant}
\begin{cases}
 & \beta_1 = \left [ \frac{L_{I_1}-L_1}{(I_1-1)Q} \right ] \text{,~for~} i=1; \\
 & \beta_i = \left [ \frac{L_{I_i}-L_{I_{i-1}}}{(I_i-I_{i-1})Q} \right ]  \text{,~for~} 2 \leq i \leq m \\
 & J_0 = \left [ \frac{L_{1}}{Q} \right ] - \beta_1  \text{,~for~} i=0; \\
 & J_i = \left [ \frac{L_{I_i}}{Q} \right ]  \text{,~for~} 1 \leq i \leq m-1,
\end{cases}
\end{equation}
where $[~]$ is the rounding operation. Again, $\beta_1$ and $J_0$ are specially treated for the first segment. A complexity reduction technique in section \ref{sub:fullORB} requires $J_{i-1}$ to be integer multiples of $\beta_i$, which can be easily achieved with operation $\left [ J_{i-1}/\beta_i \right ] \beta_i$. The segmentation method in Fig. \ref{fig:dynamic} and line parameters obtained from  Eq. \eqref{eq:quant} complete the piece-wise linear fitting and quantization.

\section{Performance and Complexity Evaluation} \label{sect:simulation}

\subsection{Decoding Performance} \label{sub:performance}

\begin{figure}[htbp]
\centerline{\includegraphics[width=0.45\textwidth]{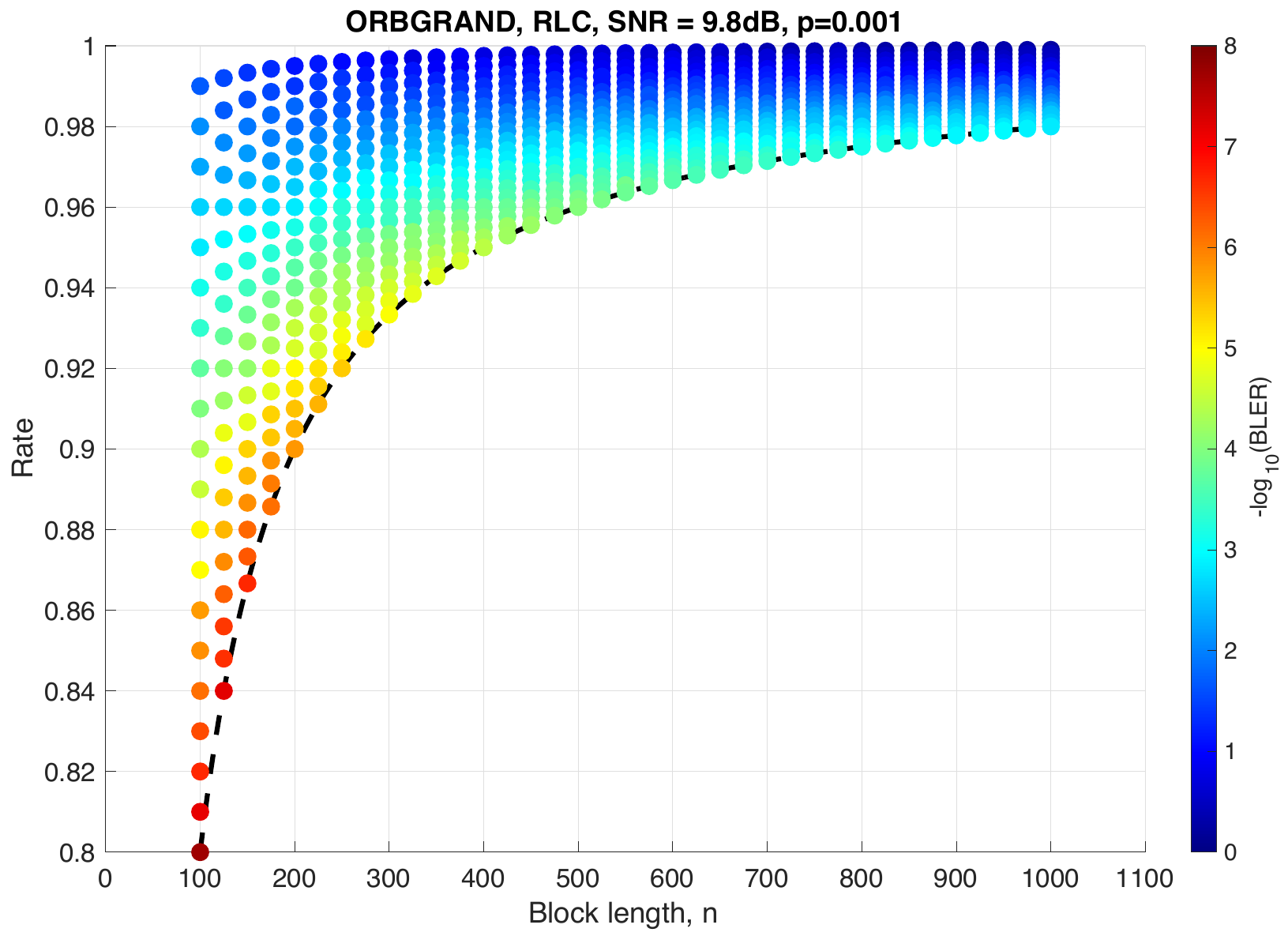}}
\caption{Decoding performance of ORBGRAND applied to RLCs of different lengths $n$ and up to $n-k=20$ redundant bits in AWGN channels using BPSK at an SNR of 9.8dB corresponding to an uncoded hard detection bit flip probability of $p=10^{-3}$.}
\label{fig:ORBGRANDrates}
\end{figure}

As explained in the introduction, an upper bound on ORBGRAND's complexity can be determined in terms of the number of parity bits in a code rather than its length or rate directly, making ORBGRAND suitable for efficient decoding of any low or moderate redundancy code. Its operational regime encompasses everything from short, low-rate codes, through higher-rate long codes. Fig. \ref{fig:ORBGRANDrates} provides a demonstration of that range of applicability by showing block error rates (BLERs), as a heat map, at a fixed SNR for Random Linear Codes (RLCs) of different code lengths and up to $20$ parity bits. While structured codes have constraints in terms of the lengths or rates at which they exist, RLCs can be constructed for any number of information and code-word bits and are known in theory to be good with high-likelihood \cite{coffey1990any}, but require a universal decoder.

To explore performance, our primary point of comparison is with CA-Polar codes as they are the state-of-the-art short, high-rate codes and they have a well-developed  soft detection decoder in CA-SCL. We first consider a CA-Polar[256, 234] code, which has 22 parity bits and uses the 11-bit CRC specified for 5G NR up-link control channels. Setting the list size to 16, which is generous in comparison to typical recommendations of $8$ \cite{xiang2019crc}, we use the CA-SCL decoder from the AFF3CT toolbox \cite{Cassagne2019a} as our performance reference. A key feature of all GRAND algorithms is that they can decode any moderate redundancy code, regardless of length or structure, and so can be used to identify the best code structures. Consequently, we also investigate BCH codes, which are known to provide excellent hard detection decoding performance but have no dedicated soft detection decoder, and CRCs, which are ubiquitously used for error detection but can be upgraded to error correction using GRAND \cite{grand-crc,LiangLiu2021}. CRCs have the desirable properties of low complexity in encoding and code-book membership checking. Finally, we also consider RLCs, whose use with GRAND variants is  being explored \cite{grand-mo, grand-crc, duffy2021ordered, Riaz21, Papadopoulou21}.

\begin{figure}[htbp]
\centerline{\includegraphics[width=0.45\textwidth]{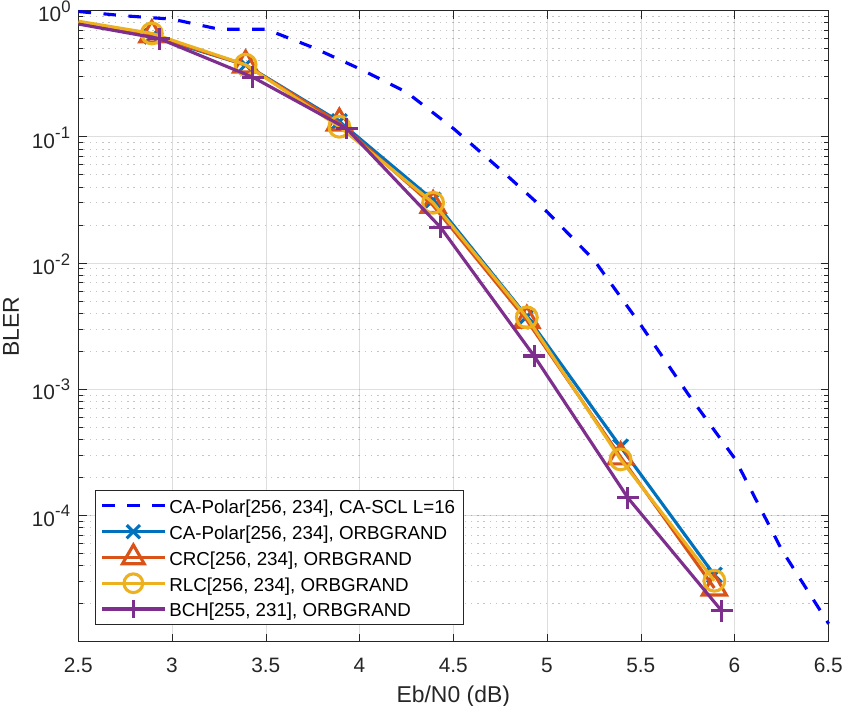}}
\caption{Decoding performance of ORBGRAND when used with CA-Polar[256, 234], CRC[256, 234], RLC[256, 234] and BCH[255,231] codes.}
\label{fig:combo_plot}
\end{figure}

Fig. \ref{fig:combo_plot} presents results using the 3-line version of ORBGRAND, as illustrated in Fig. \ref{fig:dynamic}. For this plot, ORBGRAND abandons searching and records a block error if no code-book element is identified within $5 \times 10^6$ code-book queries. As $5 \times 10^6>2^{22}$, where $22$ is the number of parity bits in the code, this threshold is sufficient to ensure ORBGRAND rarely abandons and the full error correction performance of the code is revealed. All of the codes provide near identical performance despite their distinct structures, consistent with the notion that for soft detection decoding performance is dominated by the quality of the decoder.

We next explore the impact that the number of lines and the selection of the intervals for those lines, in ORBGRAND's statistical model of the ranked reliabilities of the received bits, has on performance. To enable comparison with a well-regarded soft detection decoder, we examine three CA-Polar code configurations: CA-Polar[256, 234]; CA-Polar[512, 490]; and CA-Polar[1024, 1002]. All of those codes have 22 parity bits and employ the 11-bit CRC specified for 5G NR up-link control channels. 

\begin{figure}
\centering
\includegraphics[width=0.9\linewidth]{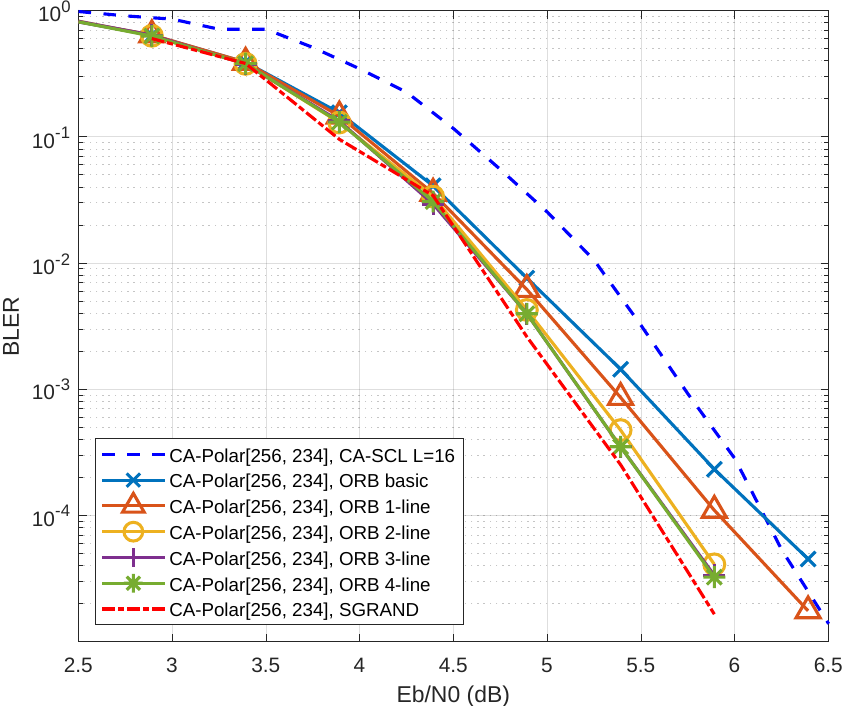}
\caption{Performance evaluation of a CA-Polar[256, 234] code as decoded with CA-SCL (list size 16) or ORBGRAND variants with normal ORBGRAND quantization.}
\label{fig:polar_256_234}
\end{figure}

Fig. \ref{fig:polar_256_234} presents simulation results for CA-Polar[256, 234], with the SGRAND results serving as the ultimate performance bound as it necessarily identifies ML decodings\cite{solomon20}. For lower values of the SNR, all ORBGRAND variants exhibit substantially better performance than CA-SCL because of the incomplete utilization of CRC bits for error correction in the CA-SCL algorithm \cite{grand-crc}. For a BLER of $10^{-4}$ or below, CA-SCL outperforms the basic variant of ORBGRAND as its model fails to  produce putative noise sequences in near-ML order at higher SNR. ORBGRAND with 1-line fitting provides an observable, but limited, improvement over the basic version, where the only difference is that the 1-line version starts from the quantized value of $\absorbLLR{1}$ instead of the origin. With the 2-line version, curvature in the low reliability region is captured, essentially eliminating any performance loss, and leaving only a small room of improvement for the 3-line version, which in turn overlaps with the 4-line version, and demonstrates close to optimal performance. 

\begin{figure}
\centering
\includegraphics[width=0.9\linewidth]{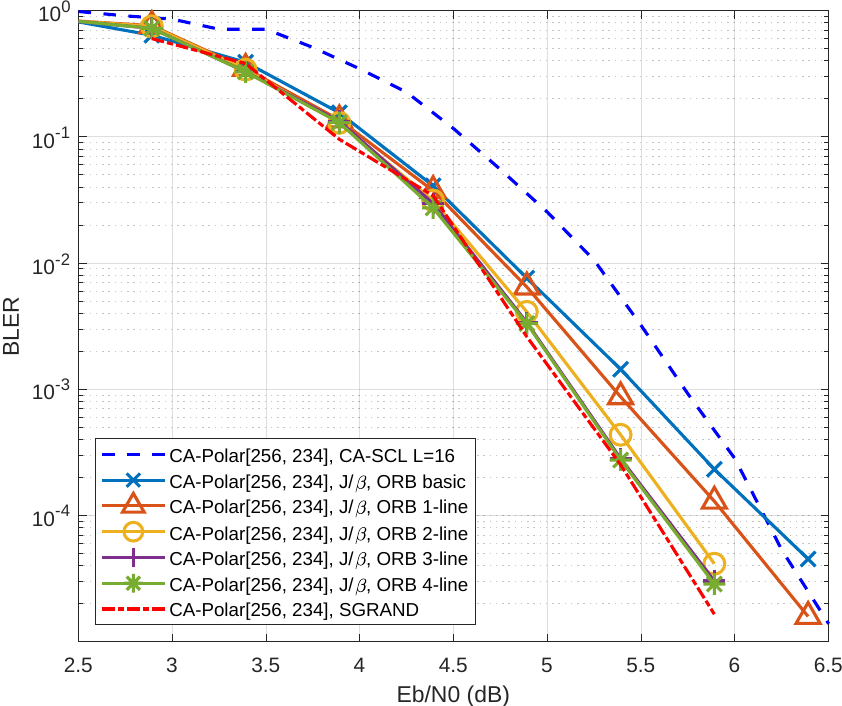}
\caption{Performance evaluation of a CA-Polar[256, 234] code as decoded with CA-SCL (list size 16) or ORBGRAND variants with $J_{i-1}$ divisible by $\beta_i$.}
\label{fig:polar_256_234_js}
\end{figure}

\begin{figure}
\centering
\includegraphics[width=0.9\linewidth]{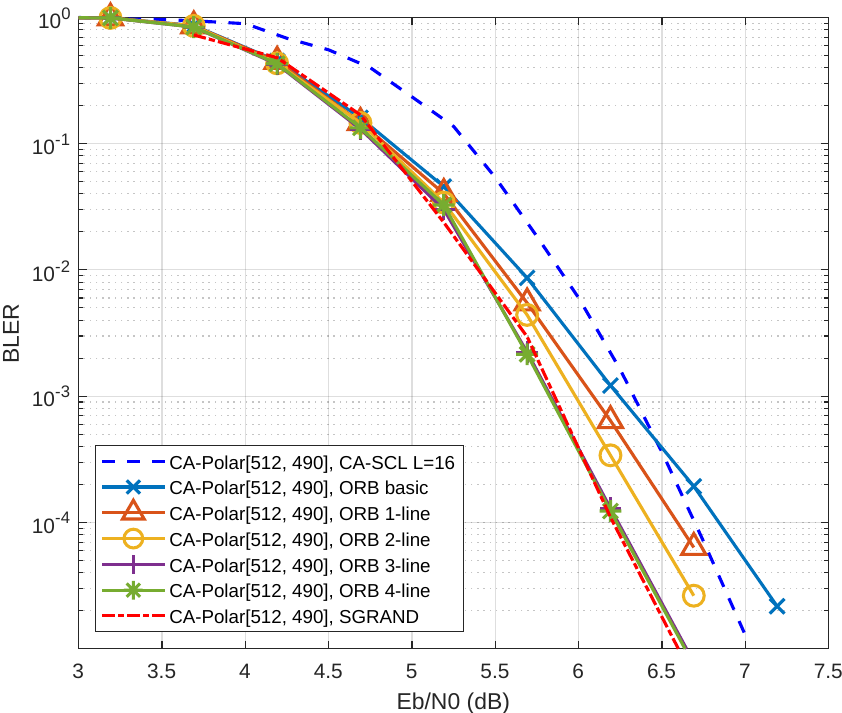}
\caption{Performance evaluation of the CA-Polar[512, 490] code decoded with CA-SCL (list size 16) or ORBGRAND variants with normal ORBGRAND quantization.}
\label{fig:polar_512_490}
\end{figure}

\begin{figure}
\centering
\includegraphics[width=0.9\linewidth]{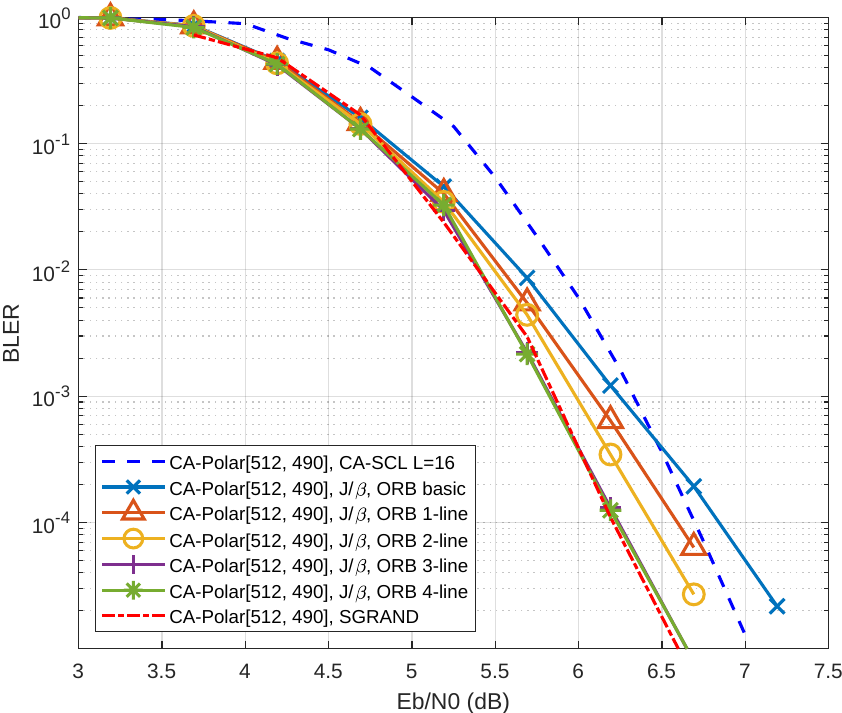}
\caption{Performance evaluation of the CA-Polar[512, 490] code decoded with  CA-SCL (list size 16) or ORBGRAND variants with $J_{i-1}$ divisible by $\beta_i$.}
\label{fig:polar_512_490_js}
\end{figure}

\begin{figure}
\centering
\includegraphics[width=0.9\linewidth]{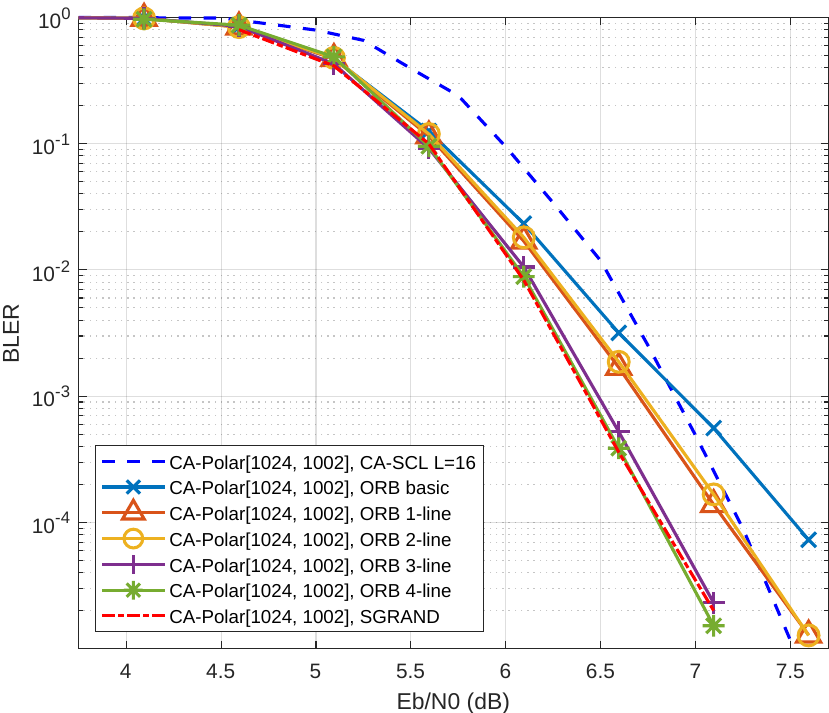}
\caption{Performance evaluation of the CA-Polar[1024, 1002] code decoded with CA-SCL (list size 16) or ORBGRAND algorithms with normal quantization.}
\label{fig:polar_1024_1002}
\end{figure}

\begin{figure}
\centering
\includegraphics[width=0.9\linewidth]{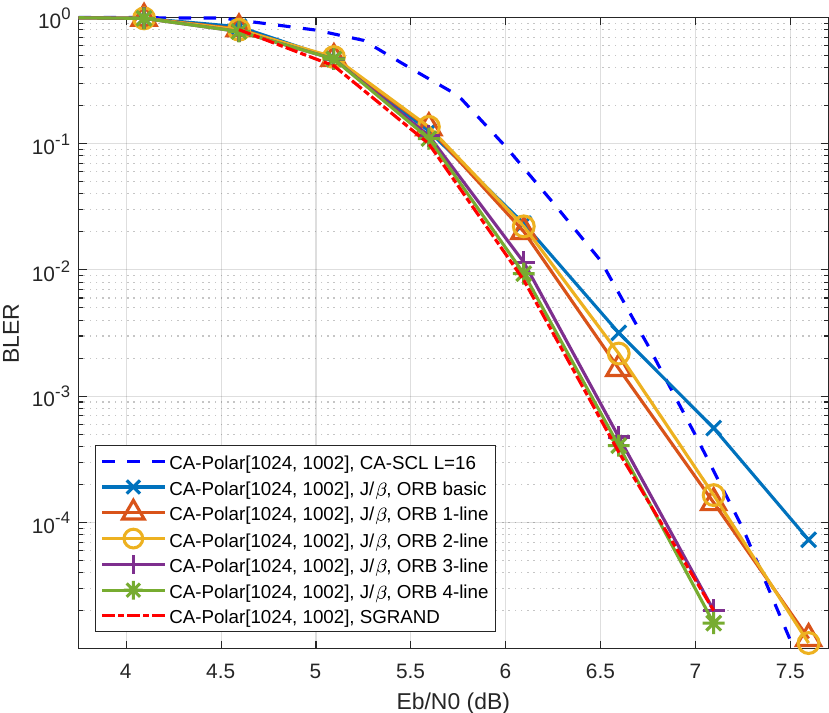}
\caption{Performance evaluation of the CA-Polar[1024, 1002] code decoded with CA-SCL(list size 16) or ORBGRAND algorithms with $J_{i-1}$ divisible by $\beta_i$.}
\label{fig:polar_1024_1002_js}
\end{figure}

Similar observations can be made for the results for the CA-Polar[512, 490] code in Fig. \ref{fig:polar_512_490} and  Fig. \ref{fig:polar_512_490_js}, and for CA-Polar[1024, 1002] code in Fig. \ref{fig:polar_1024_1002} and  Fig.\ref{fig:polar_1024_1002_js}, except that with longer block lengths, the 3-line and 4-line versions exhibit more substantial decoding improvements. Also, with the same number of parity bits, the loss of performance of the basic version occurs at a higher BLER, as shown in Fig. \ref{fig:polar_1024_1002}, where CA-SCL surpasses the basic version before a BLER of $10^{-3}$. As with the 256 bit code, the performance gap from multi-line ORBGRRAND to SGRAND is negligible, indicating its near-optimal property.

Finally, we evaluate the impact of complexity control methods, which can bring significant advantages for ORBGRAND in practical implementations, on performance. An example complexity control measure is to have $J_{i-1}$ in Eq. (\ref{eq:wu_i}) be an integer multiple of $\beta_i$. As discussed in Section \ref{sub:fullORB}, the advantage is that Algorithm \ref{alg:phm} is no longer needed, improving the efficiency of Algorithm \ref{alg:split}. As shown in Fig. \ref{fig:polar_256_234_js}, Fig. \ref{fig:polar_512_490_js} and Fig. \ref{fig:polar_1024_1002_js}, with the factor of $J_{i-1}/ \beta_i$ joined in, there is trivial change of performance between decoders with corresponding segmentation, demonstrating the robustness of ORBGRAND.

\subsection{Computational Complexity} \label{sub:complexity}
The computational complexity of any GRAND algorithm is determined by two factors, which we call the operation complexity and the code-book query number complexity. The former encompasses the computation involved in generating a single noise pattern and testing for code-book membership. The code-book query number complexity is the average number of noise pattern tested in decoding a code-word, which is SNR dependent. As established in VLSI designs for earlier GRAND variants, multiplication of these two factors forms the main complexity contribution to GRAND algorithms. 

For ORBGRAND, as well as the other soft decoding SGRAND, there is the additional step of sorting the received demodulated bits by their reliability. The study of sorting algorithms  has led to the development of numerous methods \cite{DBLP:books/aw/Knuth73}. When realized in circuits, latency, power and area are the major performance criteria and a wide collection of sorting algorithms have been efficiently implemented in ASIC and FPGA \cite{sorting19, sorting19_06, sorting21}, ranging from the simple high-latency min-max sorting algorithm \cite{sorting14} to the parallel low-latency Bitonic sorting algorithm \cite{sorting68}. For ORBGRAND, any of these approaches can be chosen depending on latency or power consumption requirements.

The noise pattern generator distinguishes variants of GRAND in terms of both decoding performance and computational complexity. The original hard detection GRAND had the simplest pattern generator, which has been efficiently implemented in hardware. 
SGRAND is at the other end of the spectrum, achieving true soft detection ML performance at the cost of a complicated pattern generation algorithm that requires large dynamic memory, making it more appropriate for performance evaluation than practical decoding. ORBGRAND, which is implementable in hardware by design, provides a range of available performance determined by the number of segments in the statistical model of reliability. As explained in Section \ref{sect:theOrb}, the core component is the Landslide algorithm, which  is highly suitable for efficient VLSI implementation, as can be understood from  the description of Fig. \ref{fig:land}. In addition to the $J/\beta$ technique for complexity control, there are further operation complexity reduction techniques proposed in \cite{An2022Th}, such as static segmentation and an efficient integer splitting algorithm. The Landslide algorithm along with those complexity control techniques means that pattern generation is not a complexity bottleneck for ORBGRAND. Consequently, for complexity evaluation we focus on the other major factor: the average number of code-book queries until a decoding is found. 


\begin{figure}[htbp]
\centerline{\includegraphics[width=0.48\textwidth]{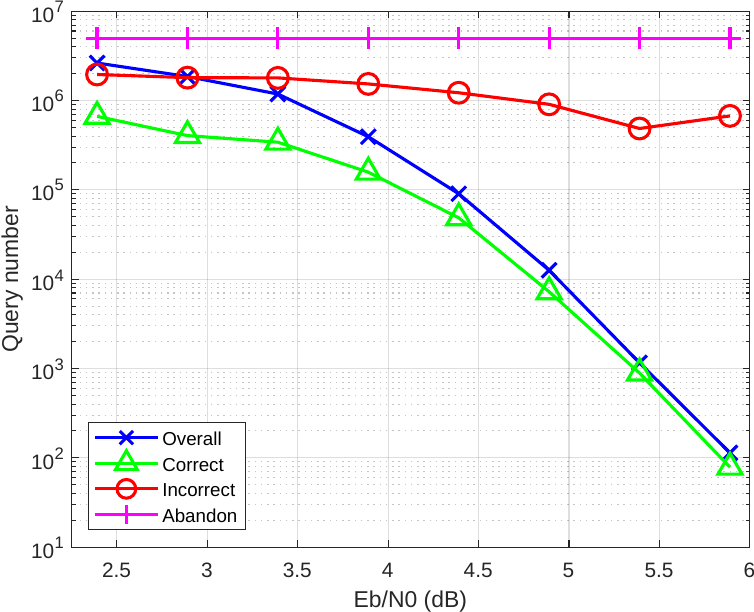}}
\caption{Complexity of 3-line ORBGRAND operating on CA-Polar[256, 234] in Fig. \ref{fig:polar_256_234} in terms of average code-book query number with: ``Overall'' for all processed code-words; ``Correct'' for correct decodings; ``Incorrect'' for incorrect decodings; and ``Abandon'' for meeting the abandonment condition prior to code-word identification.}
\label{fig:plr256_comp}
\end{figure}

We begin with a complexity investigation of 3-line ORBGRAND evaluated with a CA-Polar[256, 234] code whose BLER performance is reported in Fig. \ref{fig:polar_256_234}. Fig. \ref{fig:plr256_comp} presents the average number of code-book queries until decoding or abandonment. A core feature of all GRAND algorithms is that the number of queries they make until a decoding is found decreases quickly as channel conditions improve, with the average query number per decoding approaching the average number of correctly decoded code-words. In a standard operating regime with BLER of $10^{-3}$, the average query number is approximately $3000$ per decoding, which can be efficiently accomplished with VLSI circuits. The complexity reduces further to approximately $300$ queries per decoding as the BLER improves to $10^{-4}$, indicating low energy operation in good channel conditions. This feature suggests ORBGRAND as an appropriate candidate for an ultra low-power decoding solution.

\begin{figure*}
\centering
\begin{subfigure}{0.4\textwidth}
  \centering
  \includegraphics[width=1.0\linewidth]{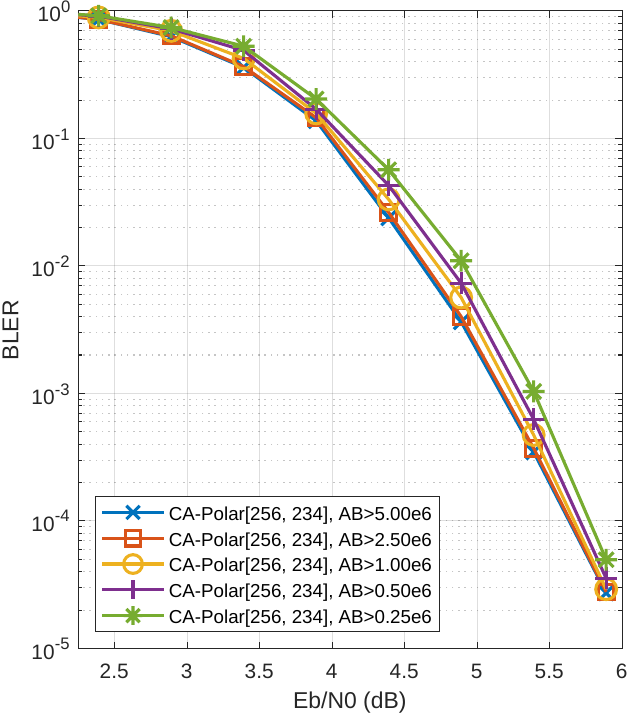}
  \caption{}
  \label{fig:complexity256_234_orb1}
\end{subfigure}%
\begin{subfigure}{0.4\textwidth}
  \centering
  \includegraphics[width=1.0\linewidth]{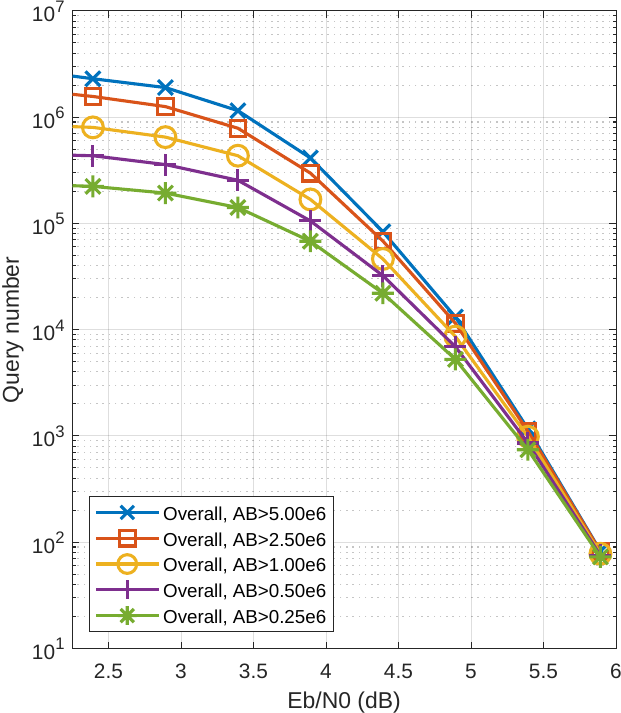}
  \caption{}
  \label{fig:complexity256_234_orb2}
\end{subfigure}
\caption{Decoding performance and average query number complexity for 3-line full ORBGRAND evaluated with the CA-Polar{[}256,234{]} code in an AWGN channel for various abandonment conditions: (a) decoding performance; (b) average query number complexity.}
\label{fig:complexity256_234_orb}
\end{figure*}

In this plot, the query number at which abandonment occurs is fixed at $5 \times 10^6$ which is greater than $2^{n-k}=2^{22}$ and so ensures optimally accurate decoding. The ``Incorrect'' curve sits under the ``Abandonment'' curve, indicating the possibility of lowering the abandonment condition and saving computation without impacting decoding performance. In practice, the abandonment threshold can be reduced with limited impact on decoding performance while saving complexity. To illustrate that feature, Fig. \ref{fig:complexity256_234_orb} presents the decoding performance and corresponding query number complexity of ORBGRAND under various abandonment conditions. The ORBGRAND configuration is identical to the 3-line scenario in Fig. \ref{fig:polar_256_234}. When the abandonment condition is reduced to $2.5\times10^6$, as suggested by the ``Incorrect'' complexity curve in Fig. \ref{fig:plr256_comp}, there is almost no performance loss but significant complexity reduction in the lower SNR region and observable complexity improvement in the operating region at a BLER of $10^{-3}$. The complexity continues to reduce as the abandonment condition lowers, incurring a slight performance degradation. At a BLER of $10^{-4}$ or below, the complexity saving is no longer significant, indicating that the choice of abandonment condition has little effect in at high SNR.

\begin{figure*}
\centering
\begin{subfigure}{0.4\textwidth}
  \centering
  \includegraphics[width=1.0\linewidth]{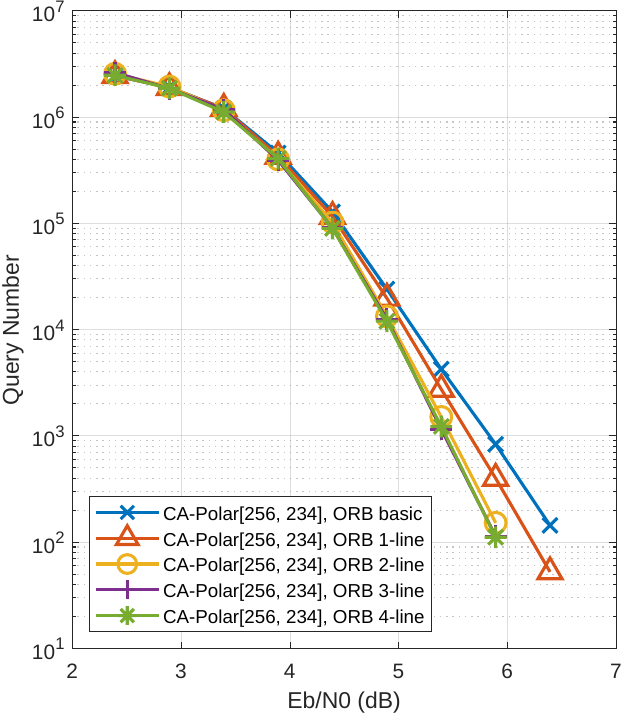}
  \caption{}
  \label{fig:complexity1}
\end{subfigure}%
\begin{subfigure}{0.4\textwidth}
  \centering
  \includegraphics[width=1.0\linewidth]{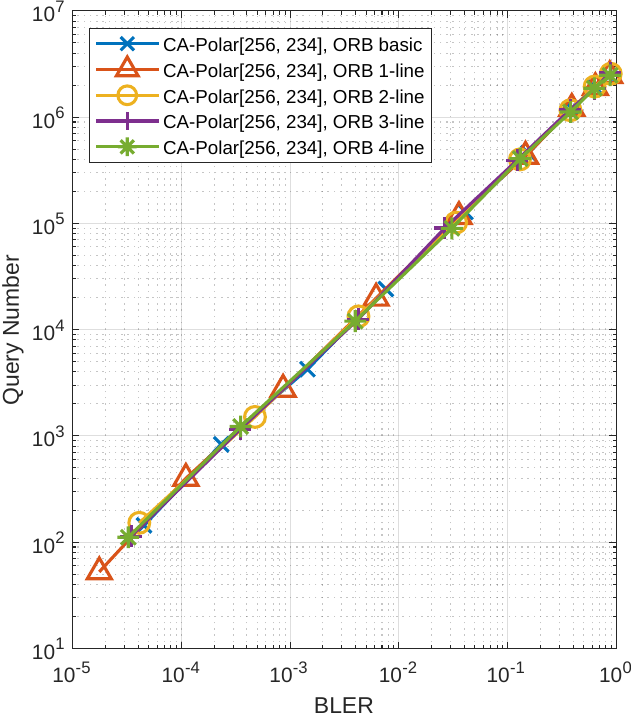}
  \caption{}
  \label{fig:complexity2}
\end{subfigure}
\caption{Average code-book query number for a CA-Polar[256, 234] decoded with ORBGRAND and an abandonment condition $5 \times 10^6$: (a) query number v.s. $E_b/N_0$; (b) query number v.s. BLER.}
\label{fig:complexity}
\end{figure*}

So far, we have been using the 3-line full ORBGRAND for the evaluation of code-book query complexity. Results in Fig. \ref{fig:polar_256_234} demonstrate that the decoding performance of ORBGRAND improves with the number of segments considered in the algorithm. Their corresponding average query number complexity is presented in Fig.\ref{fig:complexity1}. While the basic ORBGRAND has the lowest operation complexity, in the high SNR region its performance is inferior to the multi-line ORBGRAND variants and it requires more computation in terms of the average number of queries required to identify a code-word. Within the scope of full ORBGRAND, as more segments are included, resulting in a slightly increased operation complexity, better BLER performance and lower average query numbers are simultaneously achieved.

Considering both Fig. \ref{fig:polar_256_234} and Fig. \ref{fig:complexity1} at the BLER of $10^{-3}$, note that all variations of ORBGRAND have a query number of approximately $3000$, indicating the code-book number query complexity is associated with the BLER target rather than any other factor. Fig. \ref{fig:complexity2} further illustrates the relationship between query number and BLER, where the curves for all variations of ORBGRAND essentially overlap. The feature speaks to the joint enhancement of decoding performance and code-book query complexity simultaneously, justifying the return for enhanced query order designs. The near-linear curve shape in Fig. \ref{fig:complexity2} also provides a convenient tool to estimate the average overall query number for any desired BLER decoding performance. We note that this observation is consistent with the complexity analysis previously reported for the hard-detection GRAND-MO algorithm \cite{An2022TCOM}, suggesting it may be a common property to the entire family of GRAND algorithms.

\section{Discussion} \label{sect:summary}
With an abundance of new applications requiring low latency and high reliability for their operation, finding and decoding short, high-rate codes is attracting substantial attention. Old and new candidate codes along with their standard decoders have been explored and recognized to have imperfections in either the decoder or the code itself. We have introduced ORBGRAND, a practical soft detection  variant of guessing random additive noise decoding, with which it is possible to decode any moderate redundancy code with near optimal performance.

ORBGRAND offers a range of design complexities with its basic version being the simplest and requiring the least soft information. The core algorithm of the basic ORBGRAND generates integer partitions, for which we proposed the Landslide algorithm, which is suitable for efficient real-time hardware implementation. That algorithm is an essential component for the full ORBGRAND, which has higher design complexity, but can better exploit soft information at higher SNRs for additional decoding gains. Simulation results show that ORBGRAND's performance is dependent on how well the reliability curve is approximated and we proposed a piece-wise linear approximation to the reliability curve that optimizes ORBGRAND across all SNRs. 

The ORBGRAND algorithm, curve fitting techniques, and robustness to complexity improvement are established with simulations. The decoding performance is dependent on ORBGRAND's design complexity, but the 3-line version is capable of maintaining close-to-optimal performance in most scenarios. The proposed complexity control method is demonstrated to have little impact on  performance, illustrating the robustness of ORBGRAND and anticipating the potential for further complexity reduction measures to facilitate VLSI implementation. 

The practicality of the ORBGRAND algorithm is further demonstrated by assessment of its computational complexity. By design, ORBGRAND test patterns can be efficiently created, while simulated assessment of the number of code-book queries required to identify a decoding demonstrate that the approach is computational practical for moderate redundancy codes. A common feature of GRAND algorithms is confirmed from the observation that the average number of code-book queries required to identify a decoding quickly reduces as SNR improves. Controlling the abandonment condition can have a significant impact on query complexity in low SNR region, but the influence quickly fades at higher SNRs. We observe that the enhancement of ORGRAND's query order simultaneously improves both BLER performance and code-book query complexity, justifying the increased algorithmic complexity of more sophisticated noise pattern generators.

\section*{Acknowledgement}
The project or effort depicted was or is sponsored by the Defense Advanced Research Projects Agency under Grant number HR00112120008, the content of the information does not necessarily reflect the position or policy of the Government, and no official endorsement should be inferred.

\bibliographystyle{IEEEtran}
\bibliography{my_bibli}

\end{document}